\documentclass[conference,10pt]{IEEEtran}
\normalsize
\ifCLASSINFOpdf
\else
\fi

\usepackage{amsthm}
\usepackage{amsmath}
\usepackage{amssymb}
\usepackage{graphicx}
\usepackage{esint}
\usepackage{amsfonts}
\usepackage{cite}
\usepackage{balance}
\usepackage{caption}
\usepackage{subcaption}
\usepackage{epstopdf}
\usepackage{color}
\usepackage{tikz}
\usepackage{pgfplots} 
\usetikzlibrary{patterns}

\usepackage[left=1.62cm,right=1.62cm,top=1.9cm]{geometry}

\makeatletter

\theoremstyle{plain}

\makeatother

\providecommand{\theoremname}{Proposition}

\begin{document}
\title{Training Terahertz Wireless Systems to Battle \\ I/Q Imbalance}

\author{\IEEEauthorblockN{Alexandros-Apostolos A. Boulogeorgos\IEEEauthorrefmark{1}, and Angeliki Alexiou\IEEEauthorrefmark{2}
}
\IEEEauthorrefmark{2}{\footnotesize{}{}{{{{{{{Department of Electrical and Computer Engineering, University of Western Macedonia, Kozani 50100, Greece. E-mail: aboulogeorgos@uowm.gr
}}}}}}}}
\\
\IEEEauthorrefmark{2}{{\footnotesize{}{}{{{{{{{Department of Digital Systems, University of Piraeus, Piraeus 18534, Greece.
E-mail: alexiou@unipi.gr.
}}}}}}}}}

}
\maketitle

\begin{abstract}
	Due to the non-ideality of analog components, transceivers experience high levels of hardware imperfections, like in-phase and quadrature imbalance (IQI), which manifests itself as the mismatches of amplitude and phase between the I and Q branches. Unless proper mitigated, IQI has an important and negative impact on the reliability and efficiency of high-frequency and high-data-rate systems, such as terahertz wireless networks. Recognizing this, the current paper presents an intelligent transmitter (TX) and an intelligent receiver (RX) architecture that by employing machine learning (ML) methodologies is capable to fully-mitigate the impact of IQI without performing IQI coefficients estimation. They key idea lies on co-training the TX mapper's and RX demapper in order to respectively design a constellation and detection scheme that takes accounts for IQI. Two training approaches are implemented, namely: i) conventional that requires a considerable amount of data for training, and ii) a reinforcement learning based one, which demands a shorter dataset in comparison to the former. The feasibility and efficiency of the proposed architecture and training approaches are validated through respective Monte Carlo simulations. 
\end{abstract}

\section{Introduction}\label{S:Intro}

To solve the spectrum scarcity problem that fifth generation (5G) systems are destined to face, the wireless world has turned its attention to the terahertz (THz) band, in which more than $50\,\mathrm{GHz}$ contiguous bandwidth is available~\cite{Chang2022,Akyildiz2022,A:LC_CR_vs_SS,Chaccour2022,Kokkoniemi2021,Zhang2021,WP:Wireless_Thz_system_architecture_for_networks_beyond_5G,Cacciapuoti2018}. Although THz wireless systems can open the door in a number of novel applications, including holographic and extended reality, autonomous robotics, and massive self-driving vehicles, it comes with several challenges that need to be addressed, before commercialization. The most important of them is the effect of hardware imperfections~\cite{Boulogeorgos2018,Chen2019,Papasotiriou2020,Song2022}.

Scanning the technical literature, there are several published contributions that quantify the impact of hardware imperfections~\cite{ Boulogeorgos2018a,A:Effects_of_RF_impairments_In_Cascaded,Boulogeorgos2019,Abbasi2019,Sutton2021} and present possible mitigation approaches~\cite{Bassam2009,Nam2012,Gomaa2014,Ramadan2018}. For example, in~\cite{Boulogeorgos2018a}, the detrimental effect of residual hardware imperfections on the energy detection performance of full-duplex cognitive radio networks was studied. The impact of IQI on the outage performance of wireless systems that experience cascaded fading was discussed in~\cite{A:Effects_of_RF_impairments_In_Cascaded}.  In~\cite{Boulogeorgos2019}, the negative effect of hardware imperfections on the outage probability of wireless systems, which employ digital beamforming, was quantified. In~\cite{Abbasi2019}, the authors experimentally assessed the hardware imperfections performance degradation in terms of spectral efficiency in wireless systems that use hybrid beamforming. The authors of~\cite{Sutton2021} quantified the impact of IQI on the spectal efficiency in cell-free massive multiple-input multiple-output wireless~systems.  From the hardware imperfection mitigation point of view, the authors of~\cite{Bassam2009} presented an IQI coefficients estimation and IQI mitigation approaches based on digital signal processing (DSP). The mitigation efficiency of this approach depends on the accuracy of the IQI coefficients estimator. A bind IQI compensation approach that is based on the least mean square filter and recursive least square filter adaptation algorithms was presented in~\cite{Nam2012}. This approach although it achieves a high image rejection ratio (IRR), it cannot fully mitigate the impact of IQI. In~\cite{Gomaa2014}, the authors presented a data-aided IQI estimation and compensation approach for orthogonal frequency division modulation systems that isolate the IQI parameters estimation for the channel estimation. Finally, in~\cite{Ramadan2018}, an IQI pre-compensation scheme for THz wireless systems, which uses as input the estimation of the transmitter (TX) and receiver (RX) IQI coefficients was reported. The efficiency of the scheme presented in~\cite{Ramadan2018} depends on the estimation accuracy of the IQI~coefficients.           

From the aforementioned contributions, two important observations can be extracted, i.e., i) the impact of hardware imperfections and especially the IQI can significantly degrade the reliability and spectral efficiency of THz wireless systems; thus, it is necessary to utilize high-efficient hardware imperfections de-emphazation techniques; and ii) current IQI mitigation approaches cannot fully mitigate its impact. The main drawback of the nowadays mitigation approaches lies on the fact that they first perform IQI coefficients estimation using DSP, which is a relatively complex procedure, and then either use the estimated coefficients to mitigate the impact of IQI or design predistortion schemes. However, both these approaches heavily depends on the accuracy of the estimation. In order to disengage from the estimation error, we follow a different approach. We design suitable constellations and detection approaches that take into account the level of IQI, which is a specification of the transceivers, without performing any IQI estimation. In more detail, the technical contribution of this work lies on the design of an intelligent TX and RX that employ machine learning (ML) in order to respectively design their mapper and de-mapper. To train the mapper and de-mapper, we document two approaches, namely: i) conventional; and ii) reinforcement learning (RL)-based training. The former employs the Adam optimizer in order to train the TX and the stochastic gradient descent (SGD) in order to train the RX, while the latter employs SGD to train both the TX and RX. Both training approaches are designed in order to minimize the channel pollution due to information exchanging between the TX and RX. Their main difference is on the size of the training dataset that they require. RL-based training requires a dataset that is size is approximately an order of magnitude lower compared to the conventional-one. To validate the performance of the proposed approaches, respective Monte Carlo simulations that quantify the BER as well as comparison with the new radio transmission and reception scheme were performed. The results highlight the superiority of the proposed approach against the baseline concepts.

\section{System model }\label{sec:SSM}

We consider a short-range single-carrier THz wireless system that consists of a single TX and a single RX. The baseband unit of the TX is responsible of converting a block of information bits, $\mathbf{b}$, into a block of complex symbols $\mathbf{s}\in\mathbb{C}^{2^{m}\times 1}$, where $m$ stands for the number of bits loaded in each symbol. To achieve this task, a neural network (NN) maps $\mathbf{b}$ into $\mathbf{s}$, using as an additional input the system's signal-to-distortion-plus-noise-ratio (SDNR), $\gamma$. The output of the NN is a complex baseband signal that is fed to the TX radio-frequency front-end.

\begin{figure}
	\centering
	\scalebox{0.45}{

	\tikzset{every picture/.style={line width=0.75pt}} 
	
	\begin{tikzpicture}[x=0.75pt,y=0.75pt,yscale=-1,xscale=1]
		
		\draw  [fill={rgb, 255:red, 240; green, 235; blue, 235 }  ,fill opacity=0.48 ][line width=1.5]  (289.5,158.75) .. controls (289.5,139.56) and (305.06,124) .. (324.25,124) -- (565.75,124) .. controls (584.94,124) and (600.5,139.56) .. (600.5,158.75) -- (600.5,367.25) .. controls (600.5,386.44) and (584.94,402) .. (565.75,402) -- (324.25,402) .. controls (305.06,402) and (289.5,386.44) .. (289.5,367.25) -- cycle ;
		\draw  [fill={rgb, 255:red, 240; green, 235; blue, 235 }  ,fill opacity=0.48 ][line width=1.5]  (46,49.12) .. controls (46,33.04) and (59.04,20) .. (75.13,20) -- (249.87,20) .. controls (265.96,20) and (279,33.04) .. (279,49.12) -- (279,289.88) .. controls (279,305.96) and (265.96,319) .. (249.87,319) -- (75.13,319) .. controls (59.04,319) and (46,305.96) .. (46,289.88) -- cycle ;
		\draw  [fill={rgb, 255:red, 203; green, 238; blue, 230 }  ,fill opacity=0.49 ][line width=1.5]  (63,241) .. controls (63,238.24) and (65.24,236) .. (68,236) -- (128,236) .. controls (130.76,236) and (133,238.24) .. (133,241) -- (133,271) .. controls (133,273.76) and (130.76,276) .. (128,276) -- (68,276) .. controls (65.24,276) and (63,273.76) .. (63,271) -- cycle ;
		\draw [line width=1.5]    (14,256) -- (60,256) ;
		\draw [shift={(63,256)}, rotate = 180] [color={rgb, 255:red, 0; green, 0; blue, 0 }  ][line width=1.5]    (14.21,-4.28) .. controls (9.04,-1.82) and (4.3,-0.39) .. (0,0) .. controls (4.3,0.39) and (9.04,1.82) .. (14.21,4.28)   ;
		\draw [line width=1.5]    (132,256) -- (178,256) ;
		\draw [shift={(181,256)}, rotate = 180] [color={rgb, 255:red, 0; green, 0; blue, 0 }  ][line width=1.5]    (14.21,-4.28) .. controls (9.04,-1.82) and (4.3,-0.39) .. (0,0) .. controls (4.3,0.39) and (9.04,1.82) .. (14.21,4.28)   ;
		\draw  [line width=1.5]  (181,256) .. controls (181,247.72) and (187.72,241) .. (196,241) .. controls (204.28,241) and (211,247.72) .. (211,256) .. controls (211,264.28) and (204.28,271) .. (196,271) .. controls (187.72,271) and (181,264.28) .. (181,256) -- cycle ;
		\draw [line width=1.5]    (186.5,246.5) -- (205.5,265.5) ;
		\draw [line width=1.5]    (206,246) -- (186,265) ;
		\draw  [fill={rgb, 255:red, 216; green, 216; blue, 216 }  ,fill opacity=0.48 ][line width=1.5]  (115,95) .. controls (115,86.72) and (121.72,80) .. (130,80) -- (255,80) .. controls (263.28,80) and (270,86.72) .. (270,95) -- (270,185) .. controls (270,193.28) and (263.28,200) .. (255,200) -- (130,200) .. controls (121.72,200) and (115,193.28) .. (115,185) -- cycle ;
		\draw [line width=1.5]    (195,200) -- (195,238) ;
		\draw [shift={(195,241)}, rotate = 270] [color={rgb, 255:red, 0; green, 0; blue, 0 }  ][line width=1.5]    (14.21,-4.28) .. controls (9.04,-1.82) and (4.3,-0.39) .. (0,0) .. controls (4.3,0.39) and (9.04,1.82) .. (14.21,4.28)   ;
		\draw  [fill={rgb, 255:red, 204; green, 180; blue, 180 }  ,fill opacity=0.48 ][line width=1.5]  (124.63,92.17) .. controls (124.63,90.21) and (126.21,88.63) .. (128.17,88.63) -- (257.45,88.63) .. controls (259.41,88.63) and (261,90.21) .. (261,92.17) -- (261,113.45) .. controls (261,115.41) and (259.41,117) .. (257.45,117) -- (128.17,117) .. controls (126.21,117) and (124.63,115.41) .. (124.63,113.45) -- cycle ;
		\draw  [fill={rgb, 255:red, 204; green, 180; blue, 180 }  ,fill opacity=0.48 ][line width=1.5]  (125.63,131.17) .. controls (125.63,129.21) and (127.21,127.63) .. (129.17,127.63) -- (258.45,127.63) .. controls (260.41,127.63) and (262,129.21) .. (262,131.17) -- (262,152.45) .. controls (262,154.41) and (260.41,156) .. (258.45,156) -- (129.17,156) .. controls (127.21,156) and (125.63,154.41) .. (125.63,152.45) -- cycle ;
		\draw  [fill={rgb, 255:red, 204; green, 180; blue, 180 }  ,fill opacity=0.48 ][line width=1.5]  (126.55,167.17) .. controls (126.55,165.21) and (128.13,163.63) .. (130.09,163.63) -- (259.38,163.63) .. controls (261.33,163.63) and (262.92,165.21) .. (262.92,167.17) -- (262.92,188.45) .. controls (262.92,190.41) and (261.33,192) .. (259.38,192) -- (130.09,192) .. controls (128.13,192) and (126.55,190.41) .. (126.55,188.45) -- cycle ;
		\draw [line width=1.5]    (211,256) -- (290,256) ;
		\draw [shift={(293,256)}, rotate = 180] [color={rgb, 255:red, 0; green, 0; blue, 0 }  ][line width=1.5]    (14.21,-4.28) .. controls (9.04,-1.82) and (4.3,-0.39) .. (0,0) .. controls (4.3,0.39) and (9.04,1.82) .. (14.21,4.28)   ;
		\draw  [fill={rgb, 255:red, 0; green, 0; blue, 0 }  ,fill opacity=1 ][line width=1.5]  (293,256) .. controls (293,253.51) and (295.01,251.5) .. (297.5,251.5) .. controls (299.99,251.5) and (302,253.51) .. (302,256) .. controls (302,258.49) and (299.99,260.5) .. (297.5,260.5) .. controls (295.01,260.5) and (293,258.49) .. (293,256) -- cycle ;
		\draw [color={rgb, 255:red, 0; green, 0; blue, 0 }  ,draw opacity=1 ][line width=1.5]    (297.5,169) -- (297.5,272.5) ;
		\draw [color={rgb, 255:red, 0; green, 0; blue, 0 }  ,draw opacity=1 ][line width=1.5]    (297.5,260.5) -- (297.5,343) ;
		\draw [line width=1.5]    (297.5,169) -- (348,169) ;
		\draw [shift={(351,169)}, rotate = 180] [color={rgb, 255:red, 0; green, 0; blue, 0 }  ][line width=1.5]    (14.21,-4.28) .. controls (9.04,-1.82) and (4.3,-0.39) .. (0,0) .. controls (4.3,0.39) and (9.04,1.82) .. (14.21,4.28)   ;
		\draw [line width=1.5]    (297.5,343) -- (348,343) ;
		\draw [shift={(351,343)}, rotate = 180] [color={rgb, 255:red, 0; green, 0; blue, 0 }  ][line width=1.5]    (14.21,-4.28) .. controls (9.04,-1.82) and (4.3,-0.39) .. (0,0) .. controls (4.3,0.39) and (9.04,1.82) .. (14.21,4.28)   ;
		\draw  [fill={rgb, 255:red, 219; green, 243; blue, 194 }  ,fill opacity=0.5 ][line width=1.5]  (350.63,153.88) .. controls (350.63,151.18) and (352.81,149) .. (355.5,149) -- (429.13,149) .. controls (431.82,149) and (434,151.18) .. (434,153.88) -- (434,183.13) .. controls (434,185.82) and (431.82,188) .. (429.13,188) -- (355.5,188) .. controls (352.81,188) and (350.63,185.82) .. (350.63,183.13) -- cycle ;
		\draw  [fill={rgb, 255:red, 219; green, 243; blue, 194 }  ,fill opacity=0.5 ][line width=1.5]  (352.63,327.88) .. controls (352.63,325.18) and (354.81,323) .. (357.5,323) -- (431.13,323) .. controls (433.82,323) and (436,325.18) .. (436,327.88) -- (436,357.13) .. controls (436,359.82) and (433.82,362) .. (431.13,362) -- (357.5,362) .. controls (354.81,362) and (352.63,359.82) .. (352.63,357.13) -- cycle ;
		\draw [line width=1.5]    (433.5,168) -- (484,168) ;
		\draw [shift={(487,168)}, rotate = 180] [color={rgb, 255:red, 0; green, 0; blue, 0 }  ][line width=1.5]    (14.21,-4.28) .. controls (9.04,-1.82) and (4.3,-0.39) .. (0,0) .. controls (4.3,0.39) and (9.04,1.82) .. (14.21,4.28)   ;
		\draw  [fill={rgb, 255:red, 235; green, 215; blue, 176 }  ,fill opacity=0.47 ][line width=1.5]  (487,167) .. controls (487,158.72) and (493.72,152) .. (502,152) .. controls (510.28,152) and (517,158.72) .. (517,167) .. controls (517,175.28) and (510.28,182) .. (502,182) .. controls (493.72,182) and (487,175.28) .. (487,167) -- cycle ;
		\draw [fill={rgb, 255:red, 235; green, 215; blue, 176 }  ,fill opacity=0.47 ][line width=1.5]    (492.5,157.5) -- (511.5,176.5) ;
		\draw [fill={rgb, 255:red, 235; green, 215; blue, 176 }  ,fill opacity=0.47 ][line width=1.5]    (512,157) -- (492,176) ;
		
		\draw [line width=1.5]    (435.5,345) -- (486,345) ;
		\draw [shift={(489,345)}, rotate = 180] [color={rgb, 255:red, 0; green, 0; blue, 0 }  ][line width=1.5]    (14.21,-4.28) .. controls (9.04,-1.82) and (4.3,-0.39) .. (0,0) .. controls (4.3,0.39) and (9.04,1.82) .. (14.21,4.28)   ;
		\draw  [fill={rgb, 255:red, 235; green, 215; blue, 176 }  ,fill opacity=0.47 ][line width=1.5]  (489,344) .. controls (489,335.72) and (495.72,329) .. (504,329) .. controls (512.28,329) and (519,335.72) .. (519,344) .. controls (519,352.28) and (512.28,359) .. (504,359) .. controls (495.72,359) and (489,352.28) .. (489,344) -- cycle ;
		\draw [fill={rgb, 255:red, 235; green, 215; blue, 176 }  ,fill opacity=0.47 ][line width=1.5]    (494.5,334.5) -- (513.5,353.5) ;
		\draw [fill={rgb, 255:red, 235; green, 215; blue, 176 }  ,fill opacity=0.47 ][line width=1.5]    (514,334) -- (494,353) ;
		
		\draw  [fill={rgb, 255:red, 232; green, 217; blue, 233 }  ,fill opacity=0.63 ][line width=1.5]  (489,226) .. controls (489,217.72) and (495.72,211) .. (504,211) .. controls (512.28,211) and (519,217.72) .. (519,226) .. controls (519,234.28) and (512.28,241) .. (504,241) .. controls (495.72,241) and (489,234.28) .. (489,226) -- cycle ;
		\draw [line width=1.5]    (489,226) .. controls (511,210) and (497,243) .. (519,226) ;
		\draw [line width=1.5]    (504,209) -- (503.11,185) ;
		\draw [shift={(503,182)}, rotate = 87.88] [color={rgb, 255:red, 0; green, 0; blue, 0 }  ][line width=1.5]    (14.21,-4.28) .. controls (9.04,-1.82) and (4.3,-0.39) .. (0,0) .. controls (4.3,0.39) and (9.04,1.82) .. (14.21,4.28)   ;
		\draw  [fill={rgb, 255:red, 208; green, 247; blue, 153 }  ,fill opacity=0.5 ][line width=1.5]  (476,271.25) .. controls (476,268.35) and (478.35,266) .. (481.25,266) -- (522.75,266) .. controls (525.65,266) and (528,268.35) .. (528,271.25) -- (528,302.75) .. controls (528,305.65) and (525.65,308) .. (522.75,308) -- (481.25,308) .. controls (478.35,308) and (476,305.65) .. (476,302.75) -- cycle ;
		\draw [line width=1.5]    (504,241) -- (504,263) ;
		\draw [shift={(504,266)}, rotate = 270] [color={rgb, 255:red, 0; green, 0; blue, 0 }  ][line width=1.5]    (14.21,-4.28) .. controls (9.04,-1.82) and (4.3,-0.39) .. (0,0) .. controls (4.3,0.39) and (9.04,1.82) .. (14.21,4.28)   ;
		\draw [line width=1.5]    (504,304) -- (504,326) ;
		\draw [shift={(504,329)}, rotate = 270] [color={rgb, 255:red, 0; green, 0; blue, 0 }  ][line width=1.5]    (14.21,-4.28) .. controls (9.04,-1.82) and (4.3,-0.39) .. (0,0) .. controls (4.3,0.39) and (9.04,1.82) .. (14.21,4.28)   ;
		\draw [color={rgb, 255:red, 0; green, 0; blue, 0 }  ,draw opacity=1 ][line width=1.5]    (577,167) -- (517,167) ;
		\draw [line width=1.5]    (577,167) -- (577,239.5) ;
		\draw [shift={(577,242.5)}, rotate = 270] [color={rgb, 255:red, 0; green, 0; blue, 0 }  ][line width=1.5]    (14.21,-4.28) .. controls (9.04,-1.82) and (4.3,-0.39) .. (0,0) .. controls (4.3,0.39) and (9.04,1.82) .. (14.21,4.28)   ;
		\draw  [fill={rgb, 255:red, 233; green, 190; blue, 153 }  ,fill opacity=0.65 ][line width=1.5]  (562.5,257.5) .. controls (562.5,249.22) and (569.22,242.5) .. (577.5,242.5) .. controls (585.78,242.5) and (592.5,249.22) .. (592.5,257.5) .. controls (592.5,265.78) and (585.78,272.5) .. (577.5,272.5) .. controls (569.22,272.5) and (562.5,265.78) .. (562.5,257.5) -- cycle ;
		\draw [line width=1.5]    (577.5,242.5) -- (577.5,272.5) ;
		\draw [line width=1.5]    (592.5,257.5) -- (562.5,257.5) ;
		\draw [color={rgb, 255:red, 0; green, 0; blue, 0 }  ,draw opacity=1 ][line width=1.5]    (579,344) -- (519,344) ;
		\draw [line width=1.5]    (579,344) -- (577.56,275.5) ;
		\draw [shift={(577.5,272.5)}, rotate = 88.8] [color={rgb, 255:red, 0; green, 0; blue, 0 }  ][line width=1.5]    (14.21,-4.28) .. controls (9.04,-1.82) and (4.3,-0.39) .. (0,0) .. controls (4.3,0.39) and (9.04,1.82) .. (14.21,4.28)   ;
		\draw [line width=1.5]    (592.5,257.5) -- (643,257.5) ;
		\draw [shift={(646,257.5)}, rotate = 180] [color={rgb, 255:red, 0; green, 0; blue, 0 }  ][line width=1.5]    (14.21,-4.28) .. controls (9.04,-1.82) and (4.3,-0.39) .. (0,0) .. controls (4.3,0.39) and (9.04,1.82) .. (14.21,4.28)   ;
		\draw  [fill={rgb, 255:red, 240; green, 204; blue, 207 }  ,fill opacity=0.52 ][line width=1.5]  (647,206.19) .. controls (647,202.77) and (649.77,200) .. (653.19,200) -- (690.31,200) .. controls (693.73,200) and (696.5,202.77) .. (696.5,206.19) -- (696.5,317.81) .. controls (696.5,321.23) and (693.73,324) .. (690.31,324) -- (653.19,324) .. controls (649.77,324) and (647,321.23) .. (647,317.81) -- cycle ;
		\draw  [line width=1.5]  (712,231) -- (705.5,215) -- (718.5,215) -- cycle ;
		\draw [line width=1.5]    (698,248) -- (712.5,248) ;
		\draw [line width=1.5]    (712,231) -- (712,248) ;
		\draw  [line width=1.5]  (712,191) -- (705.5,175) -- (718.5,175) -- cycle ;
		\draw [line width=1.5]    (698,208) -- (712.5,208) ;
		\draw [line width=1.5]    (712,191) -- (712,208) ;
		\draw  [line width=1.5]  (712,298) -- (705.5,282) -- (718.5,282) -- cycle ;
		\draw [line width=1.5]    (698,315) -- (712.5,315) ;
		\draw [line width=1.5]    (712,298) -- (712,315) ;
		\draw [line width=1.5]    (167,45) -- (167,75) ;
		\draw [shift={(167,78)}, rotate = 270] [color={rgb, 255:red, 0; green, 0; blue, 0 }  ][line width=1.5]    (14.21,-4.28) .. controls (9.04,-1.82) and (4.3,-0.39) .. (0,0) .. controls (4.3,0.39) and (9.04,1.82) .. (14.21,4.28)   ;
		
		\draw (98,256) node   [align=left] {\begin{minipage}[lt]{37.32pt}\setlength\topsep{0pt}
				\begin{center}
					bit to\\one-hot
				\end{center}
				
		\end{minipage}};
		\draw (31,237.4) node [anchor=north west][inner sep=0.75pt]    {$\mathbf{b}$};
		\draw (144,235.4) node [anchor=north west][inner sep=0.75pt]    {$\mathbf{u}$};
		\draw (197.84,63.5) node   [align=left] {NN};
		\draw (143,94) node [anchor=north west][inner sep=0.75pt]   [align=left] {Dense layer \#1};
		\draw (143,134) node [anchor=north west][inner sep=0.75pt]   [align=left] {Dense layer \#2};
		\draw (145,169) node [anchor=north west][inner sep=0.75pt]   [align=left] {Normalization};
		\draw (199,208.4) node [anchor=north west][inner sep=0.75pt]    {$\mathcal{C}$};
		\draw (394.44,169) node    {$\mathcal{R}\{\cdot \}$};
		\draw (397.49,343) node    {$\mathcal{I}\{\cdot \}$};
		\draw (502,287) node    {$-\frac{\pi }{2}$};
		\draw (162.43,299.5) node   [align=left] {Mapper};
		\draw (438.43,384.5) node   [align=left] {I/Q modulator};
		\draw (243,237.4) node [anchor=north west][inner sep=0.75pt]    {$\mathbf{s}$};
		\draw (609,236.4) node [anchor=north west][inner sep=0.75pt]    {$\mathbf{x}$};
		\draw (652.5,300.5) node [anchor=north west][inner sep=0.75pt]  [rotate=-270] [align=left] {\begin{minipage}[lt]{57.7pt}\setlength\topsep{0pt}
				\begin{center}
					Analog\\beamformer
				\end{center}
				
		\end{minipage}};
		\draw (712.43,261) node  [font=\LARGE]  {$\vdots $};
		\draw (163,24.4) node [anchor=north west][inner sep=0.75pt]    {$\gamma $};

	\end{tikzpicture}
	}
	\caption{Intelligent TX architecture.}
	\label{Fig:TX}
\end{figure}
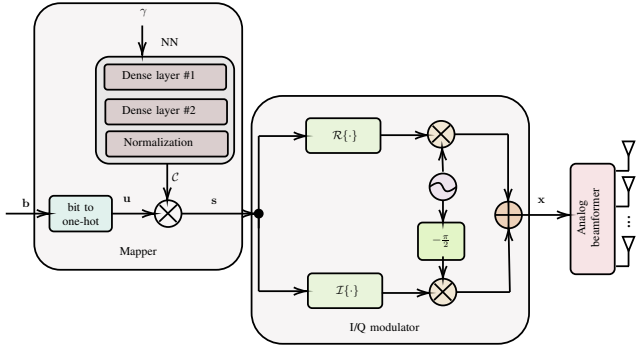
As illustrated in Fig.~\ref{Fig:TX}, the TX consists of the mapper, a conventional in-phase and quadrature (I/Q) modulator and a typical analog beamformer. The mapper has two modules, namely: i) bit to one-hot; and ii) NN. The bit to one-hot modules takes as input the bit stream of length ${m}$ and output a tuple, $\mathbf{u}$, of size $2^m$ that has all its elements equal to zero except the except the one whose index has $\mathbf{b}$ as its binary representation, which is set to one.  The NN takes as input the SDNR. It consists of two dense layer of $2^{m}+1$ units  and one normalization layer. The first dense layer uses a rectified linear unit (ReLU) for activations, while the second has linear activations. The second dense layer outputs $2^{m+1}$ elements that corresponds to the real and imaginary part of the  $2^{m}$ constellation points. The normalization layer ensures that the mean power of the generated constellation is equal to $1$.  

At the $k$-th, the output of the NN, $\mathbf{\mathcal{C}}$, is multiplied by $\mathbf{u}$ in order for the transmission symbol to be extracted, i.e.,
	$s_k = \mathbf{u}\,\mathbf{\mathcal{C}}.$
Next, $s_k$ is fed to the I/Q modulator. We assume that the I/Q modulator suffers by amplitude and phase mismatches. As a consequence, its output is distorted by I/Q imbalance (IQI).  Hence, the baseband equivalent transmitted signal at the $k-$th timeslot can be expressed as in~\cite{B:Schenk-book}
\begin{align}
	x_k = G_1\, s_k + G_2\,s_k^{*},
	\label{Eq:x_k}
\end{align} 
where $s_k$ is the $k-$th component of $\mathbf{s}$, while, according to~\cite{}, 
\begin{align}
	G_{1} = \frac{1}{2}\left(1+\epsilon_1\,e^{j\phi_1}\right)
\text{ and }
	G_{2} = \frac{1}{2}\left(1-\epsilon_1\,e^{-j\phi_1}\right).
	\label{Eq:G_2}
\end{align}
In~\eqref{Eq:G_2}, $\epsilon_1$ and $\phi_1$ respectively stand for the amplitude and phase~mismatches. 

Both the TX and the RX are assumed to employ analog beamforming in order to mitigate the high path and molecular absorption losses. In short-range THz wireless systems in which high directional antennas are employed, the line-of-sight (LoS) path plays a dominant role, since the received power by non-LoS paths is usually more than $30\,\mathrm{dB}$ lower that the received power by the LoS path. As a result, the channel coefficient can be calculated~as
$	h = \sqrt{h_p\,h_m},$
where $h_p$ stands for the geometry path gain and $h_m$ denotes the molecular absorption gain.
The geometry path gain can be evaluated, according to the free space loss expression~\cite{},~as
$	h_p = \frac{G_{t}\,G_r\,c}{\left(4\pi\right)^2 \, f^2 \, d^2},$
where $c$, $f$, and $d$ are respectively the speed of light, the carrier frequency, and the transmission distance, while $G_t$ and $G_r$ are the TX and RX antenna gains. The molecular absorption gain can be expressed~as
	$h_m = \exp\left(-\kappa(v, T, p, f)\,d\right),$
where $\kappa(v, T, p, f)$ is the absorption coefficient that depends on atmospheric conditions, like the relative humidity, $v$, the temperature, $T$, the atmospheric pressure $p$, and the carrier frequency, $f$. Note that scanning the technical literature, there are several models that can be used to model the absorption coefficient~\cite{jornet2011,Kokkoniemi2021}. In this work, for simplicity and without loss of generality, the model presented in~\cite{EuCAP2018_cr_ver7} and used in~\cite{A:Analytical_Performance_Assessment_of_THz_Wireless_Systems} was~adopted.

\begin{figure}
	\centering
	\scalebox{0.45}{
		
		\tikzset{every picture/.style={line width=0.75pt}} 
		
		\begin{tikzpicture}[x=0.75pt,y=0.75pt,yscale=-1,xscale=1]
			
			\draw  [fill={rgb, 255:red, 240; green, 235; blue, 235 }  ,fill opacity=0.48 ][line width=1.5]  (112,64.88) .. controls (112,46.72) and (126.72,32) .. (144.88,32) -- (561.13,32) .. controls (579.28,32) and (594,46.72) .. (594,64.88) -- (594,262.13) .. controls (594,280.28) and (579.28,295) .. (561.13,295) -- (144.88,295) .. controls (126.72,295) and (112,280.28) .. (112,262.13) -- cycle ;
			\draw  [fill={rgb, 255:red, 240; green, 235; blue, 235 }  ,fill opacity=0.48 ][line width=1.5]  (611,32.25) .. controls (611,19.96) and (620.96,10) .. (633.25,10) -- (766.75,10) .. controls (779.04,10) and (789,19.96) .. (789,32.25) -- (789,199.75) .. controls (789,212.04) and (779.04,222) .. (766.75,222) -- (633.25,222) .. controls (620.96,222) and (611,212.04) .. (611,199.75) -- cycle ;
			\draw  [fill={rgb, 255:red, 232; green, 217; blue, 233 }  ,fill opacity=0.63 ][line width=1.5]  (167,142) .. controls (167,133.72) and (173.72,127) .. (182,127) .. controls (190.28,127) and (197,133.72) .. (197,142) .. controls (197,150.28) and (190.28,157) .. (182,157) .. controls (173.72,157) and (167,150.28) .. (167,142) -- cycle ;
			\draw  [fill={rgb, 255:red, 240; green, 204; blue, 207 }  ,fill opacity=0.52 ][line width=1.5]  (42.57,119.69) .. controls (42.57,116.27) and (45.34,113.5) .. (48.76,113.5) -- (85.88,113.5) .. controls (89.3,113.5) and (92.07,116.27) .. (92.07,119.69) -- (92.07,231.31) .. controls (92.07,234.73) and (89.3,237.5) .. (85.88,237.5) -- (48.76,237.5) .. controls (45.34,237.5) and (42.57,234.73) .. (42.57,231.31) -- cycle ;
			\draw  [line width=1.5]  (27.57,144.5) -- (21.07,128.5) -- (34.07,128.5) -- cycle ;
			\draw [line width=1.5]    (27.57,144.5) -- (27.57,161.5) ;
			\draw [line width=1.5]    (27.57,161.5) -- (42.07,161.5) ;
			\draw  [line width=1.5]  (28.57,106.5) -- (22.07,90.5) -- (35.07,90.5) -- cycle ;
			\draw [line width=1.5]    (28.57,106.5) -- (28.57,123.5) ;
			\draw [line width=1.5]    (28.57,123.5) -- (43.07,123.5) ;
			\draw  [line width=1.5]  (27.57,214.5) -- (21.07,198.5) -- (34.07,198.5) -- cycle ;
			\draw [line width=1.5]    (27.57,214.5) -- (27.57,231.5) ;
			\draw [line width=1.5]    (27.57,231.5) -- (42.07,231.5) ;
			\draw  [fill={rgb, 255:red, 235; green, 215; blue, 176 }  ,fill opacity=0.47 ][line width=1.5]  (166,87) .. controls (166,78.72) and (172.72,72) .. (181,72) .. controls (189.28,72) and (196,78.72) .. (196,87) .. controls (196,95.28) and (189.28,102) .. (181,102) .. controls (172.72,102) and (166,95.28) .. (166,87) -- cycle ;
			\draw [fill={rgb, 255:red, 235; green, 215; blue, 176 }  ,fill opacity=0.47 ][line width=1.5]    (171.5,77.5) -- (190.5,96.5) ;
			\draw [fill={rgb, 255:red, 235; green, 215; blue, 176 }  ,fill opacity=0.47 ][line width=1.5]    (191,77) -- (171,96) ;
			
			\draw [line width=1.5]    (93,174) -- (127,174) ;
			\draw [shift={(130,174)}, rotate = 180] [color={rgb, 255:red, 0; green, 0; blue, 0 }  ][line width=1.5]    (14.21,-4.28) .. controls (9.04,-1.82) and (4.3,-0.39) .. (0,0) .. controls (4.3,0.39) and (9.04,1.82) .. (14.21,4.28)   ;
			\draw [color={rgb, 255:red, 0; green, 0; blue, 0 }  ,draw opacity=1 ][line width=1.5]    (129.5,87) -- (129.5,190.5) ;
			\draw [color={rgb, 255:red, 0; green, 0; blue, 0 }  ,draw opacity=1 ][line width=1.5]    (129.5,178.5) -- (129.5,261) ;
			\draw [line width=1.5]    (129.5,87) -- (161,87) ;
			\draw [shift={(164,87)}, rotate = 180] [color={rgb, 255:red, 0; green, 0; blue, 0 }  ][line width=1.5]    (14.21,-4.28) .. controls (9.04,-1.82) and (4.3,-0.39) .. (0,0) .. controls (4.3,0.39) and (9.04,1.82) .. (14.21,4.28)   ;
			\draw [line width=1.5]    (129.5,261) -- (159,261) ;
			\draw [shift={(162,261)}, rotate = 180] [color={rgb, 255:red, 0; green, 0; blue, 0 }  ][line width=1.5]    (14.21,-4.28) .. controls (9.04,-1.82) and (4.3,-0.39) .. (0,0) .. controls (4.3,0.39) and (9.04,1.82) .. (14.21,4.28)   ;
			\draw  [fill={rgb, 255:red, 235; green, 215; blue, 176 }  ,fill opacity=0.47 ][line width=1.5]  (164,262) .. controls (164,253.72) and (170.72,247) .. (179,247) .. controls (187.28,247) and (194,253.72) .. (194,262) .. controls (194,270.28) and (187.28,277) .. (179,277) .. controls (170.72,277) and (164,270.28) .. (164,262) -- cycle ;
			\draw [fill={rgb, 255:red, 235; green, 215; blue, 176 }  ,fill opacity=0.47 ][line width=1.5]    (169.5,252.5) -- (188.5,271.5) ;
			\draw [fill={rgb, 255:red, 235; green, 215; blue, 176 }  ,fill opacity=0.47 ][line width=1.5]    (189,252) -- (169,271) ;
			
			\draw [line width=1.5]    (167,142) .. controls (189,126) and (175,159) .. (197,142) ;
			\draw [line width=1.5]    (182,127) -- (182,105) ;
			\draw [shift={(182,102)}, rotate = 90] [color={rgb, 255:red, 0; green, 0; blue, 0 }  ][line width=1.5]    (14.21,-4.28) .. controls (9.04,-1.82) and (4.3,-0.39) .. (0,0) .. controls (4.3,0.39) and (9.04,1.82) .. (14.21,4.28)   ;
			\draw  [fill={rgb, 255:red, 208; green, 247; blue, 153 }  ,fill opacity=0.5 ][line width=1.5]  (154,187.25) .. controls (154,184.35) and (156.35,182) .. (159.25,182) -- (200.75,182) .. controls (203.65,182) and (206,184.35) .. (206,187.25) -- (206,218.75) .. controls (206,221.65) and (203.65,224) .. (200.75,224) -- (159.25,224) .. controls (156.35,224) and (154,221.65) .. (154,218.75) -- cycle ;
			\draw [line width=1.5]    (182,157) -- (182,179) ;
			\draw [shift={(182,182)}, rotate = 270] [color={rgb, 255:red, 0; green, 0; blue, 0 }  ][line width=1.5]    (14.21,-4.28) .. controls (9.04,-1.82) and (4.3,-0.39) .. (0,0) .. controls (4.3,0.39) and (9.04,1.82) .. (14.21,4.28)   ;
			\draw [line width=1.5]    (179,222) -- (179,244) ;
			\draw [shift={(179,247)}, rotate = 270] [color={rgb, 255:red, 0; green, 0; blue, 0 }  ][line width=1.5]    (14.21,-4.28) .. controls (9.04,-1.82) and (4.3,-0.39) .. (0,0) .. controls (4.3,0.39) and (9.04,1.82) .. (14.21,4.28)   ;
			\draw [line width=1.5]    (196,86) -- (221,86) ;
			\draw [shift={(224,86)}, rotate = 180] [color={rgb, 255:red, 0; green, 0; blue, 0 }  ][line width=1.5]    (14.21,-4.28) .. controls (9.04,-1.82) and (4.3,-0.39) .. (0,0) .. controls (4.3,0.39) and (9.04,1.82) .. (14.21,4.28)   ;
			\draw  [fill={rgb, 255:red, 247; green, 222; blue, 153 }  ,fill opacity=0.5 ][line width=1.5]  (226,73.25) .. controls (226,70.35) and (228.35,68) .. (231.25,68) -- (290.75,68) .. controls (293.65,68) and (296,70.35) .. (296,73.25) -- (296,104.75) .. controls (296,107.65) and (293.65,110) .. (290.75,110) -- (231.25,110) .. controls (228.35,110) and (226,107.65) .. (226,104.75) -- cycle ;
			\draw [line width=1.5]    (194,261) -- (218,261) ;
			\draw [shift={(221,261)}, rotate = 180] [color={rgb, 255:red, 0; green, 0; blue, 0 }  ][line width=1.5]    (14.21,-4.28) .. controls (9.04,-1.82) and (4.3,-0.39) .. (0,0) .. controls (4.3,0.39) and (9.04,1.82) .. (14.21,4.28)   ;
			\draw  [fill={rgb, 255:red, 247; green, 222; blue, 153 }  ,fill opacity=0.5 ][line width=1.5]  (224,247.25) .. controls (224,244.35) and (226.35,242) .. (229.25,242) -- (288.75,242) .. controls (291.65,242) and (294,244.35) .. (294,247.25) -- (294,278.75) .. controls (294,281.65) and (291.65,284) .. (288.75,284) -- (229.25,284) .. controls (226.35,284) and (224,281.65) .. (224,278.75) -- cycle ;
			\draw [line width=1.5]    (296,86) -- (321,86) ;
			\draw [shift={(324,86)}, rotate = 180] [color={rgb, 255:red, 0; green, 0; blue, 0 }  ][line width=1.5]    (14.21,-4.28) .. controls (9.04,-1.82) and (4.3,-0.39) .. (0,0) .. controls (4.3,0.39) and (9.04,1.82) .. (14.21,4.28)   ;
			\draw [line width=1.5]    (294,262) -- (319,262) ;
			\draw [shift={(322,262)}, rotate = 180] [color={rgb, 255:red, 0; green, 0; blue, 0 }  ][line width=1.5]    (14.21,-4.28) .. controls (9.04,-1.82) and (4.3,-0.39) .. (0,0) .. controls (4.3,0.39) and (9.04,1.82) .. (14.21,4.28)   ;
			\draw  [fill={rgb, 255:red, 201; green, 245; blue, 245 }  ,fill opacity=0.5 ][line width=1.5]  (325,72.25) .. controls (325,69.35) and (327.35,67) .. (330.25,67) -- (458.75,67) .. controls (461.65,67) and (464,69.35) .. (464,72.25) -- (464,103.75) .. controls (464,106.65) and (461.65,109) .. (458.75,109) -- (330.25,109) .. controls (327.35,109) and (325,106.65) .. (325,103.75) -- cycle ;
			\draw  [fill={rgb, 255:red, 201; green, 245; blue, 245 }  ,fill opacity=0.5 ][line width=1.5]  (323,246.25) .. controls (323,243.35) and (325.35,241) .. (328.25,241) -- (456.75,241) .. controls (459.65,241) and (462,243.35) .. (462,246.25) -- (462,277.75) .. controls (462,280.65) and (459.65,283) .. (456.75,283) -- (328.25,283) .. controls (325.35,283) and (323,280.65) .. (323,277.75) -- cycle ;
			\draw  [fill={rgb, 255:red, 235; green, 215; blue, 176 }  ,fill opacity=0.47 ][line width=1.5]  (488,263) .. controls (488,254.72) and (494.72,248) .. (503,248) .. controls (511.28,248) and (518,254.72) .. (518,263) .. controls (518,271.28) and (511.28,278) .. (503,278) .. controls (494.72,278) and (488,271.28) .. (488,263) -- cycle ;
			\draw [fill={rgb, 255:red, 235; green, 215; blue, 176 }  ,fill opacity=0.47 ][line width=1.5]    (493.5,253.5) -- (512.5,272.5) ;
			\draw [fill={rgb, 255:red, 235; green, 215; blue, 176 }  ,fill opacity=0.47 ][line width=1.5]    (513,253) -- (493,272) ;
			
			\draw [line width=1.5]    (503,223) -- (503,245) ;
			\draw [shift={(503,248)}, rotate = 270] [color={rgb, 255:red, 0; green, 0; blue, 0 }  ][line width=1.5]    (14.21,-4.28) .. controls (9.04,-1.82) and (4.3,-0.39) .. (0,0) .. controls (4.3,0.39) and (9.04,1.82) .. (14.21,4.28)   ;
			\draw [line width=1.5]    (462,263) -- (483,263) ;
			\draw [shift={(486,263)}, rotate = 180] [color={rgb, 255:red, 0; green, 0; blue, 0 }  ][line width=1.5]    (14.21,-4.28) .. controls (9.04,-1.82) and (4.3,-0.39) .. (0,0) .. controls (4.3,0.39) and (9.04,1.82) .. (14.21,4.28)   ;
			\draw [line width=1.5]    (464,87) -- (561.5,87) ;
			\draw [shift={(564.5,87)}, rotate = 180] [color={rgb, 255:red, 0; green, 0; blue, 0 }  ][line width=1.5]    (14.21,-4.28) .. controls (9.04,-1.82) and (4.3,-0.39) .. (0,0) .. controls (4.3,0.39) and (9.04,1.82) .. (14.21,4.28)   ;
			\draw  [fill={rgb, 255:red, 233; green, 190; blue, 153 }  ,fill opacity=0.65 ][line width=1.5]  (564.5,88.5) .. controls (564.5,80.22) and (571.22,73.5) .. (579.5,73.5) .. controls (587.78,73.5) and (594.5,80.22) .. (594.5,88.5) .. controls (594.5,96.78) and (587.78,103.5) .. (579.5,103.5) .. controls (571.22,103.5) and (564.5,96.78) .. (564.5,88.5) -- cycle ;
			\draw [line width=1.5]    (579.5,73.5) -- (579.5,103.5) ;
			\draw [line width=1.5]    (594.5,88.5) -- (564.5,88.5) ;
			\draw [color={rgb, 255:red, 0; green, 0; blue, 0 }  ,draw opacity=1 ][line width=1.5]    (580,264) -- (520,264) ;
			\draw [line width=1.5]    (580,264) -- (580,106.5) ;
			\draw [shift={(580,103.5)}, rotate = 90] [color={rgb, 255:red, 0; green, 0; blue, 0 }  ][line width=1.5]    (14.21,-4.28) .. controls (9.04,-1.82) and (4.3,-0.39) .. (0,0) .. controls (4.3,0.39) and (9.04,1.82) .. (14.21,4.28)   ;
			\draw  [fill={rgb, 255:red, 0; green, 0; blue, 0 }  ,fill opacity=1 ] (127,174) .. controls (127,172.9) and (127.9,172) .. (129,172) .. controls (130.1,172) and (131,172.9) .. (131,174) .. controls (131,175.1) and (130.1,176) .. (129,176) .. controls (127.9,176) and (127,175.1) .. (127,174) -- cycle ;
			\draw [line width=1.5]    (594.5,88.5) -- (624,88.5) ;
			\draw [shift={(627,88.5)}, rotate = 180] [color={rgb, 255:red, 0; green, 0; blue, 0 }  ][line width=1.5]    (14.21,-4.28) .. controls (9.04,-1.82) and (4.3,-0.39) .. (0,0) .. controls (4.3,0.39) and (9.04,1.82) .. (14.21,4.28)   ;
			\draw  [fill={rgb, 255:red, 216; green, 216; blue, 216 }  ,fill opacity=0.48 ][line width=1.5]  (620.31,80.81) .. controls (620.31,72.53) and (627.03,65.81) .. (635.31,65.81) -- (760.31,65.81) .. controls (768.6,65.81) and (775.31,72.53) .. (775.31,80.81) -- (775.31,170.81) .. controls (775.31,179.1) and (768.6,185.81) .. (760.31,185.81) -- (635.31,185.81) .. controls (627.03,185.81) and (620.31,179.1) .. (620.31,170.81) -- cycle ;
			\draw  [fill={rgb, 255:red, 204; green, 180; blue, 180 }  ,fill opacity=0.48 ][line width=1.5]  (628.63,76.17) .. controls (628.63,74.21) and (630.21,72.63) .. (632.17,72.63) -- (761.45,72.63) .. controls (763.41,72.63) and (765,74.21) .. (765,76.17) -- (765,97.45) .. controls (765,99.41) and (763.41,101) .. (761.45,101) -- (632.17,101) .. controls (630.21,101) and (628.63,99.41) .. (628.63,97.45) -- cycle ;
			\draw  [fill={rgb, 255:red, 204; green, 180; blue, 180 }  ,fill opacity=0.48 ][line width=1.5]  (629.63,115.17) .. controls (629.63,113.21) and (631.21,111.63) .. (633.17,111.63) -- (762.45,111.63) .. controls (764.41,111.63) and (766,113.21) .. (766,115.17) -- (766,136.45) .. controls (766,138.41) and (764.41,140) .. (762.45,140) -- (633.17,140) .. controls (631.21,140) and (629.63,138.41) .. (629.63,136.45) -- cycle ;
			\draw  [fill={rgb, 255:red, 204; green, 180; blue, 180 }  ,fill opacity=0.48 ][line width=1.5]  (630.55,151.17) .. controls (630.55,149.21) and (632.13,147.63) .. (634.09,147.63) -- (763.38,147.63) .. controls (765.33,147.63) and (766.92,149.21) .. (766.92,151.17) -- (766.92,172.45) .. controls (766.92,174.41) and (765.33,176) .. (763.38,176) -- (634.09,176) .. controls (632.13,176) and (630.55,174.41) .. (630.55,172.45) -- cycle ;
			\draw [line width=1.5]    (671,29) -- (671,59) ;
			\draw [shift={(671,62)}, rotate = 270] [color={rgb, 255:red, 0; green, 0; blue, 0 }  ][line width=1.5]    (14.21,-4.28) .. controls (9.04,-1.82) and (4.3,-0.39) .. (0,0) .. controls (4.3,0.39) and (9.04,1.82) .. (14.21,4.28)   ;
			\draw [line width=1.5]    (768.5,162.5) -- (808,162.5) ;
			\draw [shift={(811,162.5)}, rotate = 180] [color={rgb, 255:red, 0; green, 0; blue, 0 }  ][line width=1.5]    (14.21,-4.28) .. controls (9.04,-1.82) and (4.3,-0.39) .. (0,0) .. controls (4.3,0.39) and (9.04,1.82) .. (14.21,4.28)   ;
			
			\draw (48.07,214) node [anchor=north west][inner sep=0.75pt]  [rotate=-270] [align=left] {\begin{minipage}[lt]{57.7pt}\setlength\topsep{0pt}
					\begin{center}
						Analog\\beamformer
					\end{center}
					
			\end{minipage}};
			\draw (28,177.5) node  [font=\LARGE]  {$\vdots $};
			\draw (180,203) node    {$\frac{\pi }{2}$};
			\draw (261.77,88) node   [align=left] {\begin{minipage}[lt]{48.65pt}\setlength\topsep{0pt}
					\begin{center}
						Low pass \\filter
					\end{center}
					
			\end{minipage}};
			\draw (259.77,262) node   [align=left] {\begin{minipage}[lt]{48.65pt}\setlength\topsep{0pt}
					\begin{center}
						Low pass \\filter
					\end{center}
					
			\end{minipage}};
			\draw (394.5,88) node   [align=left] {\begin{minipage}[lt]{78.15pt}\setlength\topsep{0pt}
					\begin{center}
						Analog to digital \\conveter
					\end{center}
					
			\end{minipage}};
			\draw (392.5,262) node   [align=left] {\begin{minipage}[lt]{78.15pt}\setlength\topsep{0pt}
					\begin{center}
						Analog to digital \\conveter
					\end{center}
					
			\end{minipage}};
			\draw (499,199.4) node [anchor=north west][inner sep=0.75pt]    {$j$};
			\draw (701.84,47.5) node   [align=left] {NN};
			\draw (698.45,86.5) node   [align=left] {Dense layer \#1};
			\draw (698.45,126.5) node   [align=left] {Dense layer \#2};
			\draw (700.45,160.5) node   [align=left] {Dense layer \#3};
			\draw (667,8.4) node [anchor=north west][inner sep=0.75pt]    {$\gamma $};
			\draw (692.94,203) node   [align=left] {Demapper};
			\draw (594,65.4) node [anchor=north west][inner sep=0.75pt]    {$r_{k}$};
			\draw (817,146.4) node [anchor=north west][inner sep=0.75pt]    {$\hat{\mathbf{b}}$};
			\draw (362.94,48) node   [align=left] {I/Q demodulator};

		\end{tikzpicture}
	}
	\caption{Intelligent RX architecture.}
	\label{Fig:RX}
\end{figure}
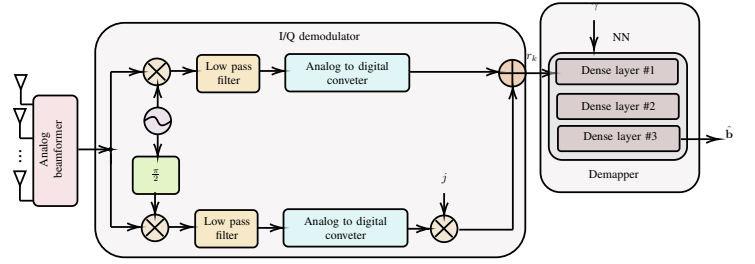

As depicted in Fig.~\ref{Fig:RX}, the RX consists of the analog beamformer, the I/Q demodulator, and the demapper. The RF received signal at the output of the RX analog beamformer is transformed to baseband signal by the I/Q demodulator. However, the I/Q demodulator suffers from IQI. Therefore,  the baseband equivalent received signal at the $k-$th timeslot can be obtained as in
\begin{align}
	r_k = K_1\,h\,x_k + K_2\,h^{*}\,x_k^{*} + K_1\, n_k + K_2\,n_k^{*},
	\label{Eq:r_k}
\end{align}
where $n_k$ is a zero-mean complex Gaussian random variable that models the additive white Gaussian noise. Moreover, $K_1$ and $K_2$ are the RX IQI coefficients that can be evaluated~as
\begin{align}
	K_1 = \frac{1}{2}\left(1+\epsilon_2\,e^{-j\,\phi_2}\right)
\text{ and }
	K_2 = \frac{1}{2}\left(1-\epsilon_2\,e{j\,\phi_2}\right),
\end{align}
with $\epsilon_2$ and $\phi_2$ respectively being the RX amplitude and phase mismatch. 
By applying~\eqref{Eq:x_k} in~\eqref{Eq:r_k}, we obtain
\begin{align}
	r_k = \xi_1\, h \, s_k + \xi_2 \, h \, s_k^{*} + K_1\,n_k + K_2 \, n_k^{*},
	\label{Eq:rk}
\end{align}
where 
$	\xi_1 = K_1\,G_1 + K_2\,G_2^{*}$
and 
$	\xi_2 = K_1\,G_2 + K_2\,G_1^{*}. $

From~\eqref{Eq:rk}, the SDNR can be written~as
\begin{align}
	\gamma= \frac{\left|\xi_1\right|^2\,h^2\,P_s}{\left|\xi_2\right|^2\,h^2\,P_s + \left(\left|K_1\right|^2 + \left|K_2\right|^2\right)\, N_o},
	\label{Eq:gamma1}
\end{align}
where  $P_s$ and $N_o$ are the transmission and noise power distributions. 

The baseband received signal is forwarder towards an NN that acts as the RX demapper. The demapper uses as an additional input the SDNR and returns an estimation of $\mathbf{b}$, i.e., ${\mathbf{\hat{b}}}$. It consists of three dense layers. The first two consists of $N$ units and employs ReLU activations. The last one has $m$ units and uses linear activation. Based on the received signal and the SDNR, the demapper estimates the probability of that each one of the symbols that belongs to $\mathbf{\mathcal{C}}$ to have been sent, and selects the one with the maximum probability.  

\section{Training the intelligent transceivers}\label{SS:Train}


\textit{Conventional training}: At the $k-$th timeslot, the TX's mapper transforms the tuple $\mathbf{u}_k$ to the symbol $s_k$ through the NN function $f_1\left(\mathbf{u};\mathbf{z}_t\right)$, where $\mathbf{z}_t$ are the TX's NN parameters. The   symbol $s_k$ is transformed to the transmitted symbol $x_k$ through the I/Q modulator that its process can be described as $f_2\left(s_k;G_1, G_2\right)$. Thus, the relationship between the input beam stream, or equivalently the input one-hot, and the transmitted signal can be described~as
$	x_k = f_2\left(f_1\left(\mathbf{u};\mathbf{z}_t\right);G_1, G_2\right),$
or, for simplicity
$	x_k = f\left(\mathbf{u}; \mathbf{z}_t, \gamma\right),$
where $f\left(\right)$ denotes the input-output relationship of the TX. 

In order to describe the channel, the conditional probability density function $p\left(y_k\left|x_k\right.\right)$ can be used. The received symbol $y_k$ is then transformed to $r_k$ through the I/Q demodulator that can be described by $f_3\left(y_k;K_1, K_2\right)$, or equivalently 
$y_k = f_3^{-1}\left(r_k;\gamma\right).$
Thus, the conditional density probability can be written~as  $p\left(f_3^{-1}\left(r_k;\gamma\right)\left|x_k\right.\right)$. Moreover, since the decoded bit stream, $\hat{\mathbf{b}}$ corresponds to the  $x_k$ and in turn to the $\mathbf{u}$ that maximizes the conditional probability, the demapper objective is to maximize $p\left(f_3^{-1}\left(r_k;\gamma\right)\left|f\left(\mathbf{u}; \mathbf{z}_t, \gamma\right)\right.\right)$. This can be achieved by minimizing a  categorical cross-entropy (CCE) function between $\mathbf{u}$ and $\mathbf{p}$, where $\mathbf{p}$ is a probability vector over all possible messages, for all the possible transmitted messages. In other words, the end-to-end training problem can be seen as the following optimization~problem:  
\begin{align}
	\begin{array}{l l }
			& \underset{{\mathbf{z}_t, \mathbf{z}_r}}{\arg\min}\,\,C_{E}\left(\mathbf{u}, \mathbf{p} \right)  \\
			\text{s.t.} &  \\
			& C_1: \quad \mathbb{E}\left[\left|\left|x\right|\right|^2\right] \leq 1
	\end{array}
\label{Eq:Cov_train}
\end{align}
where $\mathbf{z}_r$ are the RX's NN parameters and $C_{E}\left(\mathbf{u}, \mathbf{p} \right) $ denotes the  CCE function between $\mathbf{u}$ and $\mathbf{p}$ that can be defined~as
	$C_{E}\left(\mathbf{u}, \mathbf{p} \right)  = - \sum_{i=1}^{N_t} \mathbf{u}_{i}^{T} \, \log\left(\mathbf{p}_i\right),$
where $N_t$ denotes the size of the labeled dataset, $\mathbf{u}_{i}$ is the $i-$th one-hot tuple of the training data set, and $ \mathbf{p}_i$ is the corresponding probability vector. 

The optimization problem in~\eqref{Eq:Cov_train} requires both the TX and the RX to have knowledge of both $\mathbf{z}_t$ and $\mathbf{z}_r$,  in each step of the training process. This would cause an important channel pollution and the feasibility of such system would be questionable. Motivated by this the optimization problem in~\eqref{Eq:Cov_train} can break into the following two optimization problems: 
\begin{align}
	\begin{array}{l l }
		& \underset{{\mathbf{z}_t}}{\arg\min}\,\,C_{E}\left(\mathbf{u}, \mathbf{p} \right)  \\
		\text{s.t.} &  \\
		& C_1: \quad \mathbb{E}\left[\left|\left|x\right|\right|^2\right] \leq 1
	\end{array}
	\label{Eq:Cov_train1}
\end{align}
and 
\begin{align}
	\begin{array}{l l }
		& \underset{{ \mathbf{z}_r}}{\arg\min}\,\,C_{E}\left(\mathbf{u}, \mathbf{p} \right) 
	\end{array}
	\label{Eq:Cov_train2}
\end{align}
that the first one can be solved by the TX and the second by the RX with exchanging only the constellation in each training step.  The optimization problem in~\eqref{Eq:Cov_train1} is not a convex one and cannot be transformed to a convex problem. Therefore, to provide a global solution, we employ the Adam optimization method~\cite{Kingma2014}. The initial weights  are set according to the Glorot approach~\cite{Glorot2010}. On the other hand, the optimization problem in~\eqref{Eq:Cov_train2} is solved using the SGD method~\cite{Boulogeorgos2021}.      

\textit{RL-based training}: The key idea behind  RL-based training is to alternatively train the RX and the TX. In more detail, the RX uses as a the following loss function gradient: 
\begin{align}
		\bigtriangledown_{\mathbf{z}_{r}} {\mathcal{L}}&=\mathbb{E}_{\mathbf{b}, \mathbf{r}}\left[-\log\left(f_{z_R}\left(\hat{\mathbf{b}}\left|\mathbf{b}\right. \right)\right)\right].
		\label{Eq:RL_t}
\end{align}
At the TX side, the following loss function gradient is used for~training: 
\begin{align}
	\bigtriangledown_{\mathbf{z}_{t}} \hat{\mathcal{L}}&=\mathbb{E}_{\mathbf{b}, \mathbf{s},\mathbf{r}}\left[-
	\log\left(f_{z_{R}}\left(\hat{\mathbf{b}}\left|\mathbf{b}\right.\right)	\right) 
	\bigtriangledown_{\mathbf{z}_{t}} f_1\left(\mathbf{b}_1\right) 
	\right. \nonumber \\  & \times
	\left.
	\bigtriangledown_{\overline{\mathbf{b}}}\left(\log\left(\overline{\pi}_{\mathbf{s}, \mathbf{r}}\left(\mathbf{b}\right) \left|_{\overline{x}=f_{1}\left(\mathbf{b}\right)} \right. \right)\right)
	\right],
	\label{Eq:RL_2}
\end{align} 
where $f_{z_{R}}\left(\hat{\mathbf{b}}\left|\mathbf{b}\right.\right)$ denotes the estimated bits for a given block of transmission bits, and $\overline{\pi}_{\mathbf{s}, \mathbf{r}}\left(\mathbf{b}\right)$ is the estimated distribution of $\mathbf{s}$ for a given $\mathbf{b}$. Using~\eqref{Eq:RL_t} and~\eqref{Eq:RL_2}, both the TX and RX are trained  by the SGD. 

\section{Results \& Discussion}

This section focuses on presenting simulation results that highlight the effectiveness of the ML approach in mitigating the impact of IQI. The following scenario is considered.  The distance between the TX and RX is set to $10\,\mathrm{m}$ and the operation frequency is $100\,\mathrm{GHz}$. The channel bandwidth is $50\,\mathrm{MHz}$. Both the TX and RX are equipped with $30\,\mathrm{dBi}$ antennas. Standard atmospheric conditions are assumed, i.e., the relative humidity is set to $50\%$, the temperature is equal to $27\,^o\text{C}$ and the atmospheric pressure is  $101325\,\mathrm{Pa}$. Finally, $N=128$, the ML methodology with conventional training employs a dataset of $10^{5}$ data, while the ML methodology with RL training uses a dataset with $10^{4}$ data, from which $3000$ are used for fine tuning. 

\begin{figure}
	\centering
	\scalebox{0.75}{\begin{tikzpicture}
			\begin{axis}[
				ymode = log,
				xlabel={SNR (dB)},
				ylabel={BER},
				legend style={at={(1.25,0.45)},anchor=north,legend cell align=left},
				xmin = 0,
				xmax =15,
				ymin = 1e-5,
				ymax= 1.01,
				ymajorgrids=true,
				xmajorgrids=true,
				grid style=dashed,
				]
				
				\addplot[
				color=black]
				coordinates {
					(0,    0.31517)
					(0.2, 0.31206)
					(0.4, 0.31023)
					(0.6, 0.30723)
					(0.8, 0.30517)
					(1.0, 0.30296)
					(1.2, 0.30006)
					(1.4, 0.29853)
					(1.6, 0.29663)
					(1.8, 0.29379)
					(2.0, 0.29243)
					(2.2, 0.29163)
					(2.4, 0.28928)
					(2.6, 0.2874)
					(2.8, 0.28595)
					(3.0, 0.28415)
					(3.2, 0.28403)
					(3.4, 0.28153)
					(3.6, 0.28135)
					(3.8, 0.28002)
					(4.0, 0.27841)
					(4.2, 0.27729)
					(4.4, 0.27674)
					(4.6, 0.27638)
					(4.8, 0.27604)
					(5.0, 0.27646)
					(5.2, 0.27481)
					(5.4, 0.27564)
					(5.6, 0.2741)
					(5.8, 0.27268)
					(6.0, 0.27447)
					(6.2, 0.27411)
					(6.4, 0.27357)
					(6.6, 0.27345)
					(6.8, 0.27302)
					(7.0, 0.27231)
					(7.2, 0.27232)
					(7.4, 0.27225)
					(7.6, 0.27209)
					(7.8, 0.27188)
					(8.0, 0.27326)
					(8.2, 0.27188)
					(9.0, 0.27188)
					(10.0, 0.27188)
					(11.0, 0.27188)
					(12.0, 0.27188)
					(13.0, 0.27188)
					(14.0, 0.27188)
					(15.0, 0.27188)
					
				};
				\addlegendentry{\small{baseline}}
				
				\addplot[
				color=blue]
				coordinates {
					(0, 0.26891)
					(0.2, 0.26342)
					(0.4, 0.2578)
					(0.6, 0.2531)
					(0.8, 0.2468)
					(1.0, 0.24144)
					(1.2, 0.23538)
					(1.4, 0.22861)
					(1.6, 0.22305)
					(1.8, 0.2151)
					(2.0, 0.20756)
					(2.2, 0.20061)
					(2.4, 0.19074)
					(2.6, 0.18167)
					(2.8, 0.17029)
					(3.0, 0.15459)
					(3.2, 0.13886)
					(3.4, 0.11887)
					(3.6, 0.088428)
					(3.8, 0.064812)
					(4.0, 0.043707)
					(4.2, 0.023908)
					(4.4, 0.011995)
					(4.6, 0.0051116)
					(4.8, 0.0018902)
					(5.0, 5.2354E-4)
					(5.2, 1.6833E-4)
					(5.4, 2.9479E-5)
					(5.6, 4.1667E-7)
					(5.8, 2.0833E-7)
				};
				\addlegendentry{\small{ML-conv. training}}
				
				\addplot[
				color=red]
				coordinates {
					(0,    0.26415)
					(0.2, 0.26055)
					(0.4, 0.25575)
					(0.6, 0.24854)
					(0.8, 0.24416)
					(1.0, 0.23795)
					(1.2, 0.23266)
					(1.4, 0.22629)
					(1.6, 0.21981)
					(1.8, 0.21319)
					(2.0, 0.20777)
					(2.2, 0.19827)
					(2.4, 0.18935)
					(2.6, 0.17831)
					(2.8, 0.16453)
					(3.0, 0.15186)
					(3.2, 0.13412)
					(3.4, 0.11242)
					(3.6, 0.083509)
					(3.8, 0.057571)
					(4.0, 0.037867)
					(4.2, 0.019318)
					(4.4, 0.01028)
					(4.6, 0.0040087)
					(4.8, 0.0015435)
					(5.0, 4.6573E-4)
					(5.2, 1.6052E-4)
					(5.4, 2.5312E-5)
					(5.6, 3.8542E-6)
				};
				\addlegendentry{\small{ML-RL. training}}
				
				\addplot[
				only marks,
				mark = square]
				coordinates {
					(0,    0.31517)
					(0.4, 0.31023)
					(1.0, 0.30296)
					(1.4, 0.29853)
					(1.8, 0.29379)
					(2.2, 0.29163)
					(2.6, 0.2874)
					(3.0, 0.28415)
					(3.4, 0.28153)
					(3.8, 0.28002)
					(4.2, 0.27729)
					(4.6, 0.27638)
					(5.0, 0.27646)
					(5.4, 0.27564)
					(5.8, 0.27268)
					(6.2, 0.27411)
					(6.6, 0.27345)
					(7.0, 0.27231)
					(7.4, 0.27225)
					(8.0, 0.27326)
					(8.5, 0.27188)
					(9.0, 0.27188)
					(9.5, 0.27188)
					(10.0, 0.27188)
					(10.2, 0.27188)
					(10.5, 0.27188)
					(10.7, 0.27188)
					(11, 0.27188)
					(11.5, 0.27188)
					(12.0, 0.27188)
					(12.5, 0.27188)
					(13.0, 0.27188)
					(13.5, 0.27188)
					(14.0, 0.27188)
					(14.5, 0.27188)
					
				};
				\addlegendentry{\small{$\mathrm{IRR}=15\,\mathrm{dB}$}}
				
				\addplot[
				only marks,
				mark = o]
				coordinates {
					(0.2, 0.29808)
					(0.6, 0.29027)
					(1.0, 0.28274)
					(1.4, 0.27526)
					(1.8, 0.26777)
					(2.2, 0.26094)
					(2.6, 0.25271)
					(3.0, 0.2441)
					(3.4, 0.23661)
					(3.8, 0.22862)
					(4.2, 0.22009)
					(4.6, 0.21302)
					(5.0, 0.20667)
					(5.4, 0.2003)
					(5.8, 0.18957)
					(6.2, 0.18103)
					(6.6, 0.17048)
					(7.0, 0.15687)
					(7.4,0.14121)
					(7.8, 0.12093)
					(8.2, 0.099327)
					(8.6, 0.072799)
					(9.0, 0.053305)
					(9.4, 0.037201)
					(9.8, 0.023917)
					(10.2, 0.015055)
					(10.6, 0.0083706)
					(11.0, 0.0043833)
					(11.4, 0.0022128)
					(11.8, 8.5906E-4)
					(12.2, 4.9292E-4)
					(12.6, 1.3948E-4)
					(13.0, 1.1312E-4)
					(13.4, 4.8854E-5)
					(13.8, 2.9271E-5)
					(14.2, 1.5104E-5)
					(14.6, 2.0625E-5)
					(14.8, 3.7917E-5)
				};
				\addlegendentry{\small{$\mathrm{IRR}=20\,\mathrm{dB}$}}
				
				\addplot[
				only marks,
				mark = star]
				coordinates {
					(0.0, 0.31422)
					(0.2, 0.31045)
					(0.4, 0.30761)
					(0.6, 0.30266)
					(0.8, 0.29858)
					(1.0, 0.29441)
					(1.2, 0.29026)
					(1.4, 0.28747)
					(1.6, 0.28138)
					(1.8, 0.27661)
					(2.0, 0.27262)
					(2.2, 0.26743)
					(2.4, 0.26316)
					(2.6, 0.25852)
					(2.8, 0.25353)
					(3.0, 0.24849)
					(3.2, 0.24285)
					(3.4, 0.23718)
					(3.6, 0.23268)
					(3.8, 0.2251)
					(4.0, 0.22002)
					(4.2, 0.21397)
					(4.4, 0.20688)
					(4.6, 0.19822)
					(4.8, 0.19104)
					(5.0, 0.18035)
					(5.2, 0.16936)
					(5.4, 0.15587)
					(5.6, 0.13991)
					(5.8, 0.11871)
					(6.0, 0.099569)
					(6.2, 0.075025)
					(6.4, 0.053314)
					(6.6, 0.033824)
					(6.8, 0.019804)
					(7.0, 0.011023)
					(7.2, 0.0052786)
					(7.4, 0.002137)
					(7.6, 8.7604E-4)
					(7.8, 3.8062E-4)
					(8.0, 6.9896E-5)
					(8.2, 3.2083E-5)
					(8.4, 1.1458E-5)
					(8.6, 8.3333E-7)
				};
				\addlegendentry{\small{$\mathrm{IRR}=25\,\mathrm{dB}$}}
				
				\addplot[
				color=black]
				coordinates {
					(0.0, 0.30182)
					(0.2, 0.29808)
					(0.4, 0.29491)
					(0.6, 0.29027)
					(0.8, 0.28661)
					(1.0, 0.28274)
					(1.2, 0.27822)
					(1.4, 0.27526)
					(1.6, 0.27146)
					(1.8, 0.26777)
					(2.0, 0.26277)
					(2.2, 0.26094)
					(2.4, 0.2574)
					(2.6, 0.25271)
					(2.8, 0.24776)
					(3.0, 0.2441)
					(3.2, 0.24063)
					(3.4, 0.23661)
					(3.6, 0.23193)
					(3.8, 0.22862)
					(4.0, 0.22449)
					(4.2, 0.22009)
					(4.4, 0.21662)
					(4.6, 0.21302)
					(4.8, 0.20978)
					(5.0, 0.20667)
					(5.2, 0.20282)
					(5.4, 0.2003)
					(5.6, 0.19501)
					(5.8, 0.18957)
					(6.0, 0.18739)
					(6.2, 0.18103)
					(6.4, 0.17734)
					(6.6, 0.17048)
					(6.8, 0.16335)
					(7.0, 0.15687)
					(7.2, 0.14481)
					(7.4,0.14121)
					(7.6, 0.12978)
					(7.8, 0.12093)
					(8.0, 0.10792)
					(8.2, 0.099327)
					(8.4, 0.087448)
					(8.6, 0.072799)
					(8.8, 0.060787)
					(9.0, 0.053305)
					(9.2, 0.045501)
					(9.4, 0.037201)
					(9.6, 0.029926)
					(9.8, 0.023917)
					(10.0, 0.017979)
					(10.2, 0.015055)
					(10.4, 0.010675)
					(10.6, 0.0083706)
					(10.8, 0.005444)
					(11.0, 0.0043833)
					(11.2, 0.0029971)
					(11.4, 0.0022128)
					(11.6, 0.001634)
					(11.8, 8.5906E-4)
					(12.0, 7.6365E-4)
					(12.2, 4.9292E-4)
					(12.4, 3.749E-4)
					(12.6, 1.3948E-4)
					(12.8, 1.2438E-4)
					(13.0, 1.1312E-4)
					(13.2, 5.375E-5)
					(13.4, 4.8854E-5)
					(13.6, 2.2396E-5)
					(13.8, 2.9271E-5)
					(14.0, 2.1458E-5)
					(14.2, 1.5104E-5)
					(14.4, 1.9688E-5)
					(14.6, 2.0625E-5)
					(14.8, 3.7917E-5)
				};
				
				\addplot[
				black]
				coordinates {
					(0.0, 0.31422)
					(0.2, 0.31045)
					(0.4, 0.30761)
					(0.6, 0.30266)
					(0.8, 0.29858)
					(1.0, 0.29441)
					(1.2, 0.29026)
					(1.4, 0.28747)
					(1.6, 0.28138)
					(1.8, 0.27661)
					(2.0, 0.27262)
					(2.2, 0.26743)
					(2.4, 0.26316)
					(2.6, 0.25852)
					(2.8, 0.25353)
					(3.0, 0.24849)
					(3.2, 0.24285)
					(3.4, 0.23718)
					(3.6, 0.23268)
					(3.8, 0.2251)
					(4.0, 0.22002)
					(4.2, 0.21397)
					(4.4, 0.20688)
					(4.6, 0.19822)
					(4.8, 0.19104)
					(5.0, 0.18035)
					(5.2, 0.16936)
					(5.4, 0.15587)
					(5.6, 0.13991)
					(5.8, 0.11871)
					(6.0, 0.099569)
					(6.2, 0.075025)
					(6.4, 0.053314)
					(6.6, 0.033824)
					(6.8, 0.019804)
					(7.0, 0.011023)
					(7.2, 0.0052786)
					(7.4, 0.002137)
					(7.6, 8.7604E-4)
					(7.8, 3.8062E-4)
					(8.0, 6.9896E-5)
					(8.2, 3.2083E-5)
					(8.4, 1.1458E-5)
					(8.6, 8.3333E-7)
				};

				\addplot[
				only marks,
				mark = square,
				mark color=blue,
				draw = blue]
				coordinates {
					(0, 0.26891)
					(0.4, 0.2578)
					(0.8, 0.2468)
					(1.2, 0.23538)
					(1.6, 0.22305)
					(2.0, 0.20756)
					(2.4, 0.19074)
					(2.8, 0.17029)
					(3.2, 0.13886)
					(3.6, 0.088428)
					(4.0, 0.043707)
					(4.4, 0.011995)
					(4.8, 0.0018902)
					(5.2, 1.6833E-4)
					(5.6, 4.1667E-7)
				};
				
				\addplot[
				only marks,
				mark = square,
				mark color=red,
				draw = red]
				coordinates {
					(0.2, 0.26055)
					(0.6, 0.24854)
					(1.0, 0.23795)
					(1.4, 0.22629)
					(1.8, 0.21319)
					(2.2, 0.19827)
					(2.6, 0.17831)
					(3.0, 0.15186)
					(3.4, 0.11242)
					(3.8, 0.057571)
					(4.2, 0.019318)
					(4.6, 0.0040087)
					(5.0, 4.6573E-4)
					(5.4, 2.5312E-5)
				};
				
				\addplot[
				color=blue]
				coordinates {
					(0.0, 0.28389)
					(0.2, 0.2803)
					(0.4, 0.27529)
					(0.6, 0.27021)
					(0.8, 0.26503)
					(1.0, 0.26078)
					(1.2, 0.25555)
					(1.4, 0.24886)
					(1.6, 0.2447)
					(1.8, 0.23814)
					(2.0, 0.23165)
					(2.2, 0.22719)
					(2.4, 0.21887)
					(2.6, 0.2127)
					(2.8, 0.20472)
					(3.0, 0.19643)
					(3.2, 0.1885)
					(3.4, 0.17876)
					(3.6, 0.16454)
					(3.8, 0.14838)
					(4.0, 0.1299)
					(4.2, 0.11194)
					(4.4, 0.083637)
					(4.6, 0.057878)
					(4.8, 0.034826)
					(5.0, 0.020807)
					(5.2, 0.0090998)
					(5.4, 0.0038595)
					(5.6, 0.0011445)
					(5.8, 3.6375E-4)
					(6.0, 1.074E-4)
					(6.2, 1.0521E-5)
					(6.4, 2.5E-6)
				};
				
				\addplot[
				only marks,
				mark = o,
				mark color=blue,
				draw = blue]
				coordinates {
					(0.0, 0.28389)
					(0.4, 0.27529)
					(0.8, 0.26503)
					(1.2, 0.25555)
					(1.6, 0.2447)
					(2.0, 0.23165)
					(2.4, 0.21887)
					(2.8, 0.20472)
					(3.2, 0.1885)
					(3.6, 0.16454)
					(4.0, 0.1299)
					(4.4, 0.083637)
					(4.8, 0.034826)
					(5.2, 0.0090998)
					(5.6, 0.0011445)
					(6.0, 1.074E-4)
					(6.4, 2.5E-6)
				};
				
				\addplot[
				color=red]
				coordinates {
					(0.0, 0.28228)
					(0.2, 0.27926)
					(0.4, 0.27482)
					(0.6, 0.2685)
					(0.8, 0.26458)
					(1.0, 0.25871)
					(1.2, 0.25438)
					(1.4, 0.24911)
					(1.6, 0.24372)
					(1.8, 0.23819)
					(2.0, 0.23337)
					(2.2, 0.22582)
					(2.4, 0.21986)
					(2.6, 0.21202)
					(2.8, 0.20391)
					(3.0, 0.19673)
					(3.2, 0.18822)
					(3.4, 0.17728)
					(3.6, 0.16201)
					(3.8, 0.14913)
					(4.0, 0.129)
					(4.2, 0.10597)
					(4.4, 0.078357)
					(4.6, 0.052956)
					(4.8, 0.035217)
					(5.0, 0.018002)
					(5.2, 0.008064)
					(5.4, 0.0033312)
					(5.6, 0.0012808)
					(5.8, 4.5E-4)
					(6.0, 1.2521E-4)
					(6.2, 2.0E-5)
					(6.4, 9.375E-7)
				};
				
				\addplot[
				only marks,
				mark = o,
				mark color=red,
				draw = red]
				coordinates {
					(0.2, 0.27926)
					(0.6, 0.2685)
					(1.0, 0.25871)
					(1.4, 0.24911)
					(1.8, 0.23819)
					(2.2, 0.22582)
					(2.6, 0.21202)
					(3.0, 0.19673)
					(3.4, 0.17728)
					(3.8, 0.14913)
					(4.2, 0.10597)
					(4.6, 0.052956)
					(5.0, 0.018002)
					(5.4, 0.0033312)
					(5.8, 4.5E-4)
					(6.2, 2.0E-5)
				};
				
				\addplot[
				color=blue]
				coordinates {
					(0.0, 0.30913)
					(0.2, 0.30697)
					(0.4, 0.30216)
					(0.6, 0.29679)
					(0.8, 0.29319)
					(1.0, 0.28767)
					(1.2, 0.2842)
					(1.4, 0.27955)
					(1.6, 0.27455)
					(1.8, 0.27043)
					(2.0, 0.26332)
					(2.2, 0.25955)
					(2.4, 0.25404)
					(2.6, 0.24958)
					(2.8, 0.24332)
					(3.0, 0.23786)
					(3.2, 0.23265)
					(3.4, 0.22619)
					(3.6, 0.2192)
					(3.8, 0.21268)
					(4.0, 0.20497)
					(4.2, 0.19519)
					(4.4, 0.18668)
					(4.6, 0.17514)
					(4.8, 0.16383)
					(5.0, 0.14681)
					(5.2, 0.12643)
					(5.4, 0.10119)
					(5.6, 0.080922)
					(5.8, 0.049835)
					(6.0, 0.029375)
					(6.2, 0.016768)
					(6.4, 0.0075164)
					(6.6, 0.0028713)
					(6.8, 0.0010546)
					(7.0, 3.3312E-4)
					(7.2, 9.4167E-5)
					(7.4, 3.2708E-5)
					(7.6, 9.375E-7)
				};
				
				\addplot[
				only marks,
				mark = star,
				draw=blue]
				coordinates {
					(0.0, 0.30913)
					(0.2, 0.30697)
					(0.4, 0.30216)
					(0.6, 0.29679)
					(0.8, 0.29319)
					(1.0, 0.28767)
					(1.2, 0.2842)
					(1.4, 0.27955)
					(1.6, 0.27455)
					(1.8, 0.27043)
					(2.0, 0.26332)
					(2.2, 0.25955)
					(2.4, 0.25404)
					(2.6, 0.24958)
					(2.8, 0.24332)
					(3.0, 0.23786)
					(3.2, 0.23265)
					(3.4, 0.22619)
					(3.6, 0.2192)
					(3.8, 0.21268)
					(4.0, 0.20497)
					(4.2, 0.19519)
					(4.4, 0.18668)
					(4.6, 0.17514)
					(4.8, 0.16383)
					(5.0, 0.14681)
					(5.2, 0.12643)
					(5.4, 0.10119)
					(5.6, 0.080922)
					(5.8, 0.049835)
					(6.0, 0.029375)
					(6.2, 0.016768)
					(6.4, 0.0075164)
					(6.6, 0.0028713)
					(6.8, 0.0010546)
					(7.0, 3.3312E-4)
					(7.2, 9.4167E-5)
					(7.4, 3.2708E-5)
					(7.6, 9.375E-7)
				};
				
				\addplot[
				color=red]
				coordinates {
					(0.0, 0.30624)
					(0.2, 0.30257)
					(0.4, 0.29763)
					(0.6, 0.29306)
					(0.8, 0.28833)
					(1.0, 0.28451)
					(1.2, 0.27985)
					(1.4, 0.27547)
					(1.6, 0.27026)
					(1.8, 0.26669)
					(2.0, 0.26271)
					(2.2, 0.25692)
					(2.4, 0.25231)
					(2.6, 0.24604)
					(2.8, 0.24049)
					(3.0, 0.23613)
					(3.2, 0.22938)
					(3.4, 0.22312)
					(3.6, 0.21621)
					(3.8, 0.20906)
					(4.0, 0.20152)
					(4.2, 0.19265)
					(4.4, 0.18277)
					(4.6, 0.1691)
					(4.8, 0.1571)
					(5.0, 0.1411)
					(5.2, 0.11512)
					(5.4, 0.08991)
					(5.6, 0.065688)
					(5.8, 0.040688)
					(6.0, 0.024086)
					(6.2, 0.012683)
					(6.4, 0.0048109)
					(6.6, 0.0021089)
					(6.8, 6.0656E-4)
					(7.0, 1.4677E-4)
					(7.2, 2.8333E-5)
					(7.4, 1.0417E-5)
					(7.6, 2.0E-7)
				};
				
				\addplot[
				only marks,
				mark = star,
				draw = red, 
				color=red]
				coordinates {
					(0.0, 0.30624)
					(0.2, 0.30257)
					(0.4, 0.29763)
					(0.6, 0.29306)
					(0.8, 0.28833)
					(1.0, 0.28451)
					(1.2, 0.27985)
					(1.4, 0.27547)
					(1.6, 0.27026)
					(1.8, 0.26669)
					(2.0, 0.26271)
					(2.2, 0.25692)
					(2.4, 0.25231)
					(2.6, 0.24604)
					(2.8, 0.24049)
					(3.0, 0.23613)
					(3.2, 0.22938)
					(3.4, 0.22312)
					(3.6, 0.21621)
					(3.8, 0.20906)
					(4.0, 0.20152)
					(4.2, 0.19265)
					(4.4, 0.18277)
					(4.6, 0.1691)
					(4.8, 0.1571)
					(5.0, 0.1411)
					(5.2, 0.11512)
					(5.4, 0.08991)
					(5.6, 0.065688)
					(5.8, 0.040688)
					(6.0, 0.024086)
					(6.2, 0.012683)
					(6.4, 0.0048109)
					(6.6, 0.0021089)
					(6.8, 6.0656E-4)
					(7.0, 1.4677E-4)
					(7.2, 2.8333E-5)
					(7.4, 1.0417E-5)
					(7.6, 2.0E-7)
				};

			\end{axis}
			
	\end{tikzpicture}}
	\vspace{-0.2cm}
	\caption{BER vs SNR for different types of training and levels of IRR.}
	\label{Fig:BER}
\end{figure}

Figure~\ref{Fig:BER} depicts the BER as a function of the SNR for different types of training and levels of IRR, assuming $m=6$. Note that TX and RX IRRs can be respectively defined~as
	$\mathrm{IRR}_t = \frac{\left|G_1\right|^2}{\left|G_2\right|^2}$ 
\text{ and } 
	$\mathrm{IRR}_r = \frac{\left|K_1\right|^2}{\left|K_2\right|^2}. $
For this scenario, we assumed that $\mathrm{IRR}_t=\mathrm{IRR}_r=\mathrm{IRR}$. 
Moreover, we used as a baseline a new radio modulation and coding scheme that employs low-density parity-check code with code rate equal to $0.5$ and 64 quadrature amplitude modulation (QAM). As expected, for  given $\mathrm{IRR}$, transmission and reception scheme as well as training methodology, as the SNR increases,  the BER decreases. 
Moreover, from this figure, it becomes evident that for fixed $\mathrm{IRR}$ and SNR, the ML methodologies outperform the baseline approach. For instance, for $\mathrm{IRR}=20\,\mathrm{dB}$ and $\mathrm{SNR}=5\,\mathrm{dB}$, using the baseline approach would lead to  BER that are respectively equal to  $0.2067$. On the other hand, for the same $\mathrm{IRR}$ and $\mathrm{SNR}$ values, if ML with conventional training was employed, the BER would be respectively equal to  $0.0208$, while if ML with RL training was used, the BER would be respectively equal t$0.018$. This indicative example reveals the capability of ML to enable the use of low-cost transceivers, which otherwise could not be used, while significantly reducing the required SNR and in turn the energy consumption of the THz wireless system. Another observation from the previous example is that ML with RL training outperforms ML with conventional training. Likewise, we observe that for the baseline, for a fixed SNR, as $\mathrm{IRR}$ increases, i.e., the severity of IQI becomes less detrimental, the  BER decreases. On the other hand, when the ML methodologies are employed, for a given SNR, as $\mathrm{IRR}$ increases, the BER increases. The reason behind this is that the ML methodologies output a new constellation that exploits the diversity that can be gained by the TX and RX IQI~\cite{A:IQSC}. 

\begin{figure}
	\centering
	\scalebox{0.75}{\begin{tikzpicture}
			\begin{axis}[
				ymode = log,
				xlabel={SNR (dB)},
				ylabel={BER},
				legend style={at={(1.25,0.45)},anchor=north,legend cell align=left},
				xmin = 0,
				xmax =15,
				ymin = 1e-5,
				ymax= 1.01,
				ymajorgrids=true,
				xmajorgrids=true,
				grid style=dashed,
				]
				
				\addplot[
				color=black]
				coordinates {
					(0,    0.31517)
					(0.4, 0.31023)
					(1.0, 0.30296)
					(1.4, 0.29853)
					(1.8, 0.29379)
					(2.2, 0.29163)
					(2.6, 0.2874)
					(3.0, 0.28415)
					(3.4, 0.28153)
					(3.8, 0.28002)
					(4.2, 0.27729)
					(4.6, 0.27638)
					(5.0, 0.27646)
					(5.4, 0.27564)
					(5.8, 0.27268)
					(6.2, 0.27411)
					(6.6, 0.27345)
					(7.0, 0.27231)
					(7.4, 0.27225)
					(8.0, 0.27326)
					(8.5, 0.27188)
					(9.0, 0.27188)
					(9.5, 0.27188)
					(10.0, 0.27188)
					(10.2, 0.27188)
					(10.5, 0.27188)
					(10.7, 0.27188)
					(11, 0.27188)
					(11.5, 0.27188)
					(12.0, 0.27188)
					(12.5, 0.27188)
					(13.0, 0.27188)
					(13.5, 0.27188)
					(14.0, 0.27188)
					(14.5, 0.27188)
					
				};
				\addlegendentry{\small{baseline}}
				
				\addplot[
				color=blue]
				coordinates {
					(0, 0.26891)
					(0.2, 0.26342)
					(0.4, 0.2578)
					(0.6, 0.2531)
					(0.8, 0.2468)
					(1.0, 0.24144)
					(1.2, 0.23538)
					(1.4, 0.22861)
					(1.6, 0.22305)
					(1.8, 0.2151)
					(2.0, 0.20756)
					(2.2, 0.20061)
					(2.4, 0.19074)
					(2.6, 0.18167)
					(2.8, 0.17029)
					(3.0, 0.15459)
					(3.2, 0.13886)
					(3.4, 0.11887)
					(3.6, 0.088428)
					(3.8, 0.064812)
					(4.0, 0.043707)
					(4.2, 0.023908)
					(4.4, 0.011995)
					(4.6, 0.0051116)
					(4.8, 0.0018902)
					(5.0, 5.2354E-4)
					(5.2, 1.6833E-4)
					(5.4, 2.9479E-5)
					(5.6, 4.1667E-7)
					(5.8, 2.0833E-7)
				};
				\addlegendentry{\small{ML-conv. training}}
				
				\addplot[
				color=red]
				coordinates {
					(0,    0.26415)
					(0.2, 0.26055)
					(0.4, 0.25575)
					(0.6, 0.24854)
					(0.8, 0.24416)
					(1.0, 0.23795)
					(1.2, 0.23266)
					(1.4, 0.22629)
					(1.6, 0.21981)
					(1.8, 0.21319)
					(2.0, 0.20777)
					(2.2, 0.19827)
					(2.4, 0.18935)
					(2.6, 0.17831)
					(2.8, 0.16453)
					(3.0, 0.15186)
					(3.2, 0.13412)
					(3.4, 0.11242)
					(3.6, 0.083509)
					(3.8, 0.057571)
					(4.0, 0.037867)
					(4.2, 0.019318)
					(4.4, 0.01028)
					(4.6, 0.0040087)
					(4.8, 0.0015435)
					(5.0, 4.6573E-4)
					(5.2, 1.6052E-4)
					(5.4, 2.5312E-5)
					(5.6, 3.8542E-6)
				};
				\addlegendentry{\small{ML-RL. training}}
				
				\addplot[
				only marks,
				mark = square]
				coordinates {
					(0,    0.31517)
					(0.4, 0.31023)
					(1.0, 0.30296)
					(1.4, 0.29853)
					(1.8, 0.29379)
					(2.2, 0.29163)
					(2.6, 0.2874)
					(3.0, 0.28415)
					(3.4, 0.28153)
					(3.8, 0.28002)
					(4.2, 0.27729)
					(4.6, 0.27638)
					(5.0, 0.27646)
					(5.4, 0.27564)
					(5.8, 0.27268)
					(6.2, 0.27411)
					(6.6, 0.27345)
					(7.0, 0.27231)
					(7.4, 0.27225)
					(8.0, 0.27326)
					(8.5, 0.27188)
					(9.0, 0.27188)
					(9.5, 0.27188)
					(10.0, 0.27188)
					(10.2, 0.27188)
					(10.5, 0.27188)
					(10.7, 0.27188)
					(11, 0.27188)
					(11.5, 0.27188)
					(12.0, 0.27188)
					(12.5, 0.27188)
					(13.0, 0.27188)
					(13.5, 0.27188)
					(14.0, 0.27188)
					(14.5, 0.27188)
					
				};
				\addlegendentry{\small{$m=6$}}
				
				\addplot[
				only marks,
				mark = o,
				draw=black]
				coordinates {
					(0.0, 0.24008)
					(0.2, 0.23431)
					(0.4, 0.23065)
					(0.6, 0.22676)
					(0.8, 0.223)
					(1.0,0.21765)
					(1.2, 0.21356)
					(1.4, 0.20805)
					(1.6, 0.20424)
					(1.8, 0.19941)
					(2.0, 0.19658)
					(2.2, 0.19129)
					(2.4, 0.18485)
					(2.6, 0.17735)
					(2.8, 0.16826)
					(3.0, 0.1598)
					(3.2, 0.15125)
					(3.4,0.1409)
					(3.6, 0.12493)
					(3.8, 0.11032)
					(4.0, 0.093133)
					(4.2, 0.078624)
					(4.4, 0.061856)
					(4.6, 0.049866)
					(4.8, 0.036653)
					(5.0, 0.026172)
					(5.2, 0.019249)
					(5.4, 0.013779)
					(5.6, 0.0084326)
					(5.8, 0.0049436)
					(6.0, 0.0027307)
					(6.2, 0.0018142)
					(6.4, 0.0010658)
					(6.6, 7.8531E-4)
					(6.8, 3.0563E-4)
					(7.0, 1.0396E-4)
					(7.2, 7.1146E-5)
					(7.4, 1.6979E-5)
					(7.8, 2.0833E-7)
				};
				\addlegendentry{\small{$m=4$}}
				
				\addplot[
				only marks,
				mark = star,
				draw=black]
				coordinates {
					(0.0, 0.001543)
					(0.2, 4.2177E-4)
					(0.4, 2.8125E-5)
					(0.6, 9.6875E-6)
					(0.8, 2.0833E-7)
				};
				\addlegendentry{\small{$m=2$}}

				\addplot[
				only marks,
				mark = square,
				draw = blue, 
				color=blue]
				coordinates {
					(0, 0.26891)
					(0.4, 0.2578)
					(0.8, 0.2468)
					(1.2, 0.23538)
					(1.6, 0.22305)
					(2.0, 0.20756)
					(2.4, 0.19074)
					(2.8, 0.17029)
					(3.2, 0.13886)
					(3.6, 0.088428)
					(4.0, 0.043707)
					(4.4, 0.011995)
					(4.8, 0.0018902)
					(5.2, 1.6833E-4)
					(5.6, 4.1667E-7)
				};
				
				\addplot[
				only marks,
				mark = square,
				draw= red,
				color=red]
				coordinates {
					(0.2, 0.26055)
					(0.6, 0.24854)
					(1.0, 0.23795)
					(1.4, 0.22629)
					(1.8, 0.21319)
					(2.2, 0.19827)
					(2.6, 0.17831)
					(3.0, 0.15186)
					(3.4, 0.11242)
					(3.8, 0.057571)
					(4.2, 0.019318)
					(4.6, 0.0040087)
					(5.0, 4.6573E-4)
					(5.4, 2.5312E-5)
				};
				
				\addplot[
				color=black]
				coordinates {
					(0.0, 0.24008)
					(0.2, 0.23431)
					(0.4, 0.23065)
					(0.6, 0.22676)
					(0.8, 0.223)
					(1.0,0.21765)
					(1.2, 0.21356)
					(1.4, 0.20805)
					(1.6, 0.20424)
					(1.8, 0.19941)
					(2.0, 0.19658)
					(2.2, 0.19129)
					(2.4, 0.18485)
					(2.6, 0.17735)
					(2.8, 0.16826)
					(3.0, 0.1598)
					(3.2, 0.15125)
					(3.4,0.1409)
					(3.6, 0.12493)
					(3.8, 0.11032)
					(4.0, 0.093133)
					(4.2, 0.078624)
					(4.4, 0.061856)
					(4.6, 0.049866)
					(4.8, 0.036653)
					(5.0, 0.026172)
					(5.2, 0.019249)
					(5.4, 0.013779)
					(5.6, 0.0084326)
					(5.8, 0.0049436)
					(6.0, 0.0027307)
					(6.2, 0.0018142)
					(6.4, 0.0010658)
					(6.6, 7.8531E-4)
					(6.8, 3.0563E-4)
					(7.0, 1.0396E-4)
					(7.2, 7.1146E-5)
					(7.4, 1.6979E-5)
					(7.8, 2.0833E-7)
				};
				
				\addplot[
				color=blue]
				coordinates {
					(0.0, 0.2038)
					(0.2, 0.19384)
					(0.4, 0.18473)
					(0.6, 0.17206)
					(0.8, 0.15916)
					(1.0, 0.13429)
					(1.2, 0.11674)
					(1.4, 0.084181)
					(1.6, 0.0549)
					(1.8, 0.031818)
					(2.0, 0.013636)
					(2.2, 0.0057182)
					(2.4, 0.0017342)
					(2.6, 5.5063E-4)
					(2.8, 7.0417E-5)
					(3.0, 4.4271E-5)
					(4.0, 4.2708E-6)
				};
				
				\addplot[
				only marks,
				mark = o,
				draw = blue,
				color=blue]
				coordinates {
					(0.0, 0.2038)
					(0.2, 0.19384)
					(0.4, 0.18473)
					(0.6, 0.17206)
					(0.8, 0.15916)
					(1.0, 0.13429)
					(1.2, 0.11674)
					(1.4, 0.084181)
					(1.6, 0.0549)
					(1.8, 0.031818)
					(2.0, 0.013636)
					(2.2, 0.0057182)
					(2.4, 0.0017342)
					(2.6, 5.5063E-4)
					(2.8, 7.0417E-5)
					(3.0, 4.4271E-5)
					(4.0, 4.2708E-6)
				};
				
				\addplot[
				color=red]
				coordinates {
					(0.0, 0.20446)
					(0.2, 0.19416)
					(0.4, 0.18218)
					(0.6, 0.17296)
					(0.8, 0.15767)
					(1.0, 0.13808)
					(1.2, 0.11481)
					(1.4, 0.088885)
					(1.6, 0.058135)
					(1.8, 0.03166)
					(2.0, 0.015229)
					(2.2, 0.0063754)
					(2.4, 0.0020415)
					(2.6, 4.551E-4)
					(2.8, 1.6385E-4)
					(3.0, 4.375E-6)
				};
				\addplot[
				only marks,
				mark = o,
				draw = red,
				color=red]
				coordinates {
					(0.0, 0.20446)
					(0.2, 0.19416)
					(0.4, 0.18218)
					(0.6, 0.17296)
					(0.8, 0.15767)
					(1.0, 0.13808)
					(1.2, 0.11481)
					(1.4, 0.088885)
					(1.6, 0.058135)
					(1.8, 0.03166)
					(2.0, 0.015229)
					(2.2, 0.0063754)
					(2.4, 0.0020415)
					(2.6, 4.551E-4)
					(2.8, 1.6385E-4)
					(3.0, 4.375E-6)
				};
				
				\addplot[
				color=black]
				coordinates {
					(0.0, 0.001543)
					(0.2, 4.2177E-4)
					(0.4, 2.8125E-5)
					(0.6, 9.6875E-6)
					(0.8, 2.0833E-7)
				};
				
				\addplot[
				color=blue]
				coordinates {
					(0.0, 0.0020417)
					(0.2, 3.8052E-4)
					(0.4, 5.6458E-5)
					(0.6, 5.3125E-6)
					(0.8, 2.9167E-6)
				};
				
				\addplot[
				only marks,
				mark = star,
				draw = blue, 
				color=blue]
				coordinates {
					(0.0, 0.0020417)
					(0.2, 3.8052E-4)
					(0.4, 5.6458E-5)
					(0.6, 5.3125E-6)
					(0.8, 2.9167E-6)
				};
				
				\addplot[
				color=red]
				coordinates {
					(0.0, 0.0017317)
					(0.2, 3.4958E-4)
					(0.4, 5.2292E-5)
					(0.6, 1.1042E-5)
					(0.8, 1.0E-6)
				};
				
				\addplot[
				only marks,
				mark = star,
				draw = red,
				color=red]
				coordinates {
					(0.0, 0.0017317)
					(0.2, 3.4958E-4)
					(0.4, 5.2292E-5)
					(0.6, 1.1042E-5)
					(0.8, 1.0E-6)
				};

			\end{axis}
			
	\end{tikzpicture}}
	\vspace{-0.2cm}
	\caption{BER vs SNR for different types of training and levels of IRR.}
	\label{Fig:BER2}
\end{figure}

Figure~\ref{Fig:BER2} illustrates the error performance of the ML-empowered THz wireless system. In particular, the  BER is plotted against the SNR for different values of $m$, assuming that the $\mathrm{IRR}$ is equal to $15\,\mathrm{dB}$.  Again, for both the baseline and ML-empowered schemes as well as a fixed $m$, as the SNR increases,the BER decreases. For example, for the baseline, for $m=4$, as SNR increases from $4$ to $7$, the BER changes from $9.31\times 10^{-2}$ to $1.04\times 10^{-4}$. Moreover, for given transmission and reception scheme as well as SNR, as $m$ increases, the complexity of the transmission constellation increases; thus, the BER  increases.  Likewise, we observe that, for medium and high values of $m$, where the constellations have higher complexity, the ML-empowered methodologies outperform the baseline approaches in terms of error performance. For instance,  for $m=4$ and SNR equals to $2\,\mathrm{dB}$, the achievable  BER is  equal to  $0.19$, when the baseline transmission scheme is used. On the other hand, for the same $m$ and SNR, the  BER is equal to  $1.36\times 10^{-2}$, when the ML with conventional training methodology is employed, and equal to $1.52\times 10^{-2}$, when the ML with RL training methodology is used. Moreover, for $m=6$ and a SNR that is equal to $5\,\mathrm{dB}$, the  BER is equal to $1$ and $0.18$, in the case that the baseline scheme is employed. For the same $m$ and SNR, if the ML methodology with conventional training was employed, the  BER would be  $0.14$, while, if the ML methodology with RL training was used, the  BER would be r $0.14$. From these examples, it becomes evident that ML with RL training is the optimal selection for high $m$ THz wireless systems, while, for medium $m$, the optimal selection is the ML with conventional training. Finally, from this figure, we observe that the baseline outperforms in term of error rate both the ML-empowered methodologies, in the low $m$ region.  

\section{Conclusions}

In this paper, we presented an intelligent TX and an intelligent RX architecture that enables compensation of the impact of IQI without IQI coefficient estimation. The idea was based on automatically co-designing constellation and detection schemes that maps bits to symbols and received signals to bits. To achieve this, a ML approach was followed that was build upon two NN, one at the TX and the other at the RX. Two methods was used to train the NNs, i.e., i) conventional and ii) RL-based training. The latter require a considerable lower number of training data than the former and achieves similar performance in terms of BER. Both the training approaches were designed in order to minimize the channel pollution through training data exchange between the TX and RX. To validate the feasibility and quantify the efficiency of the proposed concept, Monte Carlo simulations were performed. The results revealed that for low constellation order, baseline approaches achieve the minimum  BER. On the other hand, for medium constellation order, the ML with conventional training achieved the minimum BER. Moreover, for high constellation order, the ML with RL-based training outperformed both the ML with conventional training and the baseline.  Finally, it was highlighted that for a fixed  BER requirement, the required SNR when employing ML is always lower or equal than the required SNR with baseline approaches. This indicates that intelligent transceivers may lead to greener THz wireless systems.  

\balance
\bibliographystyle{IEEEtran}
\bibliography{IEEEabrv,References}

\end{document}